\documentclass[journal]{IEEEtran}

\usepackage{amsmath,amssymb,amsfonts}
\usepackage{mathtools}
\usepackage{graphicx}
\usepackage[caption=false,font=footnotesize]{subfig}
\usepackage{booktabs}
\usepackage{array}
\usepackage{multirow}
\usepackage{subfig}
\usepackage{url}
\usepackage{siunitx}
\usepackage{xcolor}
\usepackage[ruled,vlined,linesnumbered]{algorithm2e}
\usepackage{cite}

\SetKwInput{Require}{Require}
\SetKwInput{Ensure}{Ensure}
\SetKwComment{Comment}{$\triangleright$\ }{}

\begin{document}

\title{Drift-Aware RL-based Wavelet Denoising for Network-Traffic Anomaly Detection}

\author{Priyalakshmi Sheela~\IEEEmembership{Member,~IEEE}, and Indrakshi Dey~\IEEEmembership{Senior Member,~IEEE}
\thanks{P. Sheela and I. Dey are with the Walton Institute, Department of Computing, South East Technological University, Waterford, Ireland. (Email: priyalakshmi.sheela@setu.ie, indrakshi.dey@setu.ie)}

\thanks{This work is supported in part by Taighde Eireann - Research Ireland under Grant 13/RC/2077 P2.

\textit{This work has been submitted to the IEEE for possible publication.
Copyright may be transferred without notice, after which this version
may no longer be accessible.}

}
}

\maketitle

\begin{abstract}
Traffic-utilisation measurements for network monitoring are corrupted by additive noise and \emph{statistical drift}: time-dependent change in the signal's mean, variance, distributional shape, or tail behaviour. Static wavelet denoising, calibrated under stationary independent and identically distributed (i.i.d.) Gaussian assumptions, becomes mismatched under drift and, at moderate-to-high signal-to-noise ratio (SNR), over-suppresses useful structure and degrades monitoring decisions. We propose a drift-aware framework treating adaptive wavelet denoising as a preprocessing layer optimised for two tasks: \emph{anomaly
detection}, recovering the multi-scale transient load bursts that noise and drift obscure, and \emph{capacity estimation}, recovering the operational required capacity $C_{95}$ ($95$th percentile of utilisation). Because localised bursts are multi-scale structure a wavelet preserves but a low-pass filter removes, detection discriminates denoiser families. A four-detector gate (Page-Hinkley, variance-ratio, Jensen-Shannon, Anderson-Darling) determines when to invoke a learned policy, and a Proximal Policy Optimization agent selects a per-window wavelet configuration over a mixed discrete-continuous action space. Unlike prior work, the reward is downstream task utility, not reconstruction fidelity. The denoiser is benchmarked, per drift type and input SNR, against a low-pass moving-average filter, VisuShrink, SureShrink, BayesShrink, and a Wiener filter. Defining the anomaly target on the clean signal and the drift gate on the corruption keeps both stages non-circular.

\end{abstract}


\section{Introduction}
\IEEEPARstart{N}{etwork} traffic monitoring builds on fine-grained
utilisation measurements to drive operational decisions~\cite{dey2026degeneracy}. Two monitoring functions
built on this telemetry are \emph{anomaly detection}, flagging genuine, often
short-lived load excursions so the operator can act, and \emph{capacity
estimation}, quantifying the load headroom needed for provisioning. The telemetry
feeding these functions is, however, a noisy and frequently non-stationary
observation of the underlying load. Two distinct corruptions matter: stationary
background noise, and \emph{statistical drift}, time-dependent change in the
statistical properties of the signal or its noise process, affecting the mean,
variance, distribution, temporal correlation, or tail behaviour.

Wavelet denoising is a well-established signal-enhancement method that exploits the multiresolution structure of a signal to separate informative components from noise-dominated fluctuations~\cite{mallat1989theory}. The wavelet-shrinkage methods of Donoho and Johnstone showed that nonlinear coefficient thresholding recovers signals effectively under additive noise~\cite{donoho1994ideal,donoho1995noising}, with threshold calibration derived under independent and identically distributed (i.i.d.) Gaussian noise of constant variance~\cite{donoho1995adapting,chang2000adaptive,donoho1998minimax}. In real traffic-utilisation measurements this stationarity assumption frequently fails. Because threshold-based denoising is parameter-sensitive, a threshold chosen under one regime becomes inappropriate when the variance, mean level, spectral content, or tail probability changes: a variance increase leaves residual noise or, conversely, a fixed threshold suppresses meaningful coefficients; a mean shift biases the approximation coefficients; structural change redistributes energy across scales; and tail amplification causes impulsive samples to be mistaken for high-frequency signal.

A second, less obvious failure is specific to the decision tasks. At moderate-to-high SNR, the signal is already relatively clean, and an aggressive fixed-threshold rule removes genuine dynamics: it can erase the very short, transient load excursions an anomaly detector must catch, and it biases the upper-tail statistic a capacity estimator reports. Denoising tuned only for reconstruction fidelity can therefore \emph{hurt} the downstream task. We therefore optimise the denoiser for the task objective, detection accuracy and capacity-estimation error, rather than for a fidelity surrogate.

We address both problems by casting adaptive wavelet denoising as a sequential decision problem and solving it with reinforcement learning (RL)~\cite{sutton2018reinforcement}. Unlike static rules or supervised learning, RL adapts through trial-and-error interaction guided by a reward that reflects task utility. We formulate the problem as a Markov Decision Process (MDP) in which, at each window, an agent observes the signal/noise context together with drift evidence and selects a wavelet-denoising configuration. The reward is the gain in downstream task performance, anomaly-detection accuracy and capacity-estimation accuracy, of the fixed downstream operators applied to the denoised stream. The detection target is defined on the clean signal, separately from the drift gate.

The contributions of this paper are summarised as follows.
\begin{itemize}
\item We formulate adaptive wavelet denoising as a task-aware preprocessing layer for
network monitoring under statistical drift. The proposed framework supports two
downstream tasks, namely multi-scale transient anomaly detection and capacity
estimation, and optimizes the denoising policy using downstream task utility rather
than reconstruction fidelity alone.
\item We design a complementary four-detector drift gate, Page-Hinkley, variance-ratio, Jensen-Shannon divergence, and Anderson-Darling, fused by logical OR, whose binary flags and continuous statistics are embedded in the RL state so that the policy can condition on drift presence, type, and intensity.
\item We define the anomaly task to be both non-circular with the gate and \emph{denoiser-discriminating}: anomalies are genuine multi-scale transient bursts in the \emph{clean} load, detected by a residual transient-energy operator. Because a low-pass filter removes such bursts along with noise, this task separates wavelet denoising from naive smoothing; capacity is the upper-tail statistic $C_{95}$,the 95th percentile of utilisation.
\item We integrate a task-utility reward (detection accuracy and capacity-error reduction, via fixed parameter-light operators) with PPO over a mixed discrete--continuous wavelet action space (mother wavelet, decomposition depth, thresholding function and strategy, continuous threshold gain and per-scale shaping).
\item We benchmark against a low-pass moving-average filter, classical state-of-the-art shrinkage rules (VisuShrink, SureShrink, BayesShrink), and a Wiener filter under a common protocol, reporting the detection receiver operating characteristic (ROC) and its area under the curve (AUC), and the capacity error, per drift type and per input SNR, with the denoising $\Delta$SNR as an intermediate diagnostic.
\end{itemize}

The remainder of the paper is organised as follows. Section~\ref{sec:sysmodel} describes the observation model, dataset construction, drift synthesis, and the two downstream tasks. Section~\ref{sec:drift_detection} presents the drift-detection layer. Section~\ref{sec:rl} develops the special case of MDP, the task-utility reward, the fixed downstream operators, and the PPO solution. Section~\ref{sec:results} reports the experimental results, and Section~\ref{sec:conclusion} concludes.

\section{System Model and Problem Formulation}
\label{sec:sysmodel}

\subsection{Observation Model and Data Augmentation}
The signal of interest is a traffic-utilisation measurement stream (dominant period $\approx 48$ samples) obtained at a network monitoring point. The observation model is
\begin{equation}
    y_{t} \;=\; x_{t} \;+\; n_{t} \;+\; d_{t},
\label{eq:obs_model}
\end{equation}
where $x_{t}$ is the clean reference load, $n_{t}$ is additive white Gaussian noise (AWGN), and $d_{t}$ is the drift-induced perturbation. A single reference trajectory is insufficient to train an RL policy. To enlarge the clean-signal distribution while preserving temporal structure, we augment the references using neural style transfer (NST)~\cite{iwana2021empirical,gatys2016image}, which is intended to preserve spectral and morphological characteristics (high correlation with the original) so that learned wavelet configurations remain meaningful. 

\subsection{Drift Synthesis}

Let $x_t^{(i)}$, $t=1,2,\ldots,T$, denote the $i$-th reference utilisation trajectory, where $T=701$ is the number of temporal samples in each complete trajectory. The purpose of the drift-synthesis stage is not merely to increase the number of samples, but to generate diverse, statistically interpretable non-stationary regimes for training the succeeding reinforcement learning (RL) framework. In the proposed framework, the detector first identifies whether a current window departs from the reference regime. If drift is detected, the RL agent selects an adaptive wavelet-denoising configuration rather than using a fixed baseline denoising rule. Therefore, the synthesized dataset must expose the detector and the RL agent to a sufficiently broad range of drift locations, durations, strengths, and noise levels.

For each reference trajectory $x_t^{(i)}$, an AWGN-corrupted no-drift baseline is first generated as $y_t^{(i,0)} = x_t^{(i)} + n_t^{(i)}$, with $n_t^{(i)} \sim \mathcal{N}(0,\sigma_n^2)$. The AWGN variance is determined by the prescribed input signal-to-noise ratio (SNR) and reference signal power as $\sigma_n^2 = P_x/10^{\mathrm{SNR}_{\mathrm{dB}}/10}$, where $P_x = \tfrac{1}{T}\sum_{t=1}^{T}(x_t^{(i)})^2$ is the reference signal power.
The SNR values used for dataset construction are from severe, moderate, and mild background-noise regimes. Low-SNR cases are included to evaluate robustness under strong channel or measurement noise, whereas high-SNR cases test whether subtle drift signatures remain detectable when the background noise is weak.

Each drift instance is injected into a contiguous temporal segment $\mathcal{T}_d = \{t_s,t_s+1,\ldots,t_e\}\subset\{1,2,\ldots,T\}$,
where $t_s$ and $t_e$ denote the drift start and end indices, respectively. These indices are determined using a normalized onset parameter $\alpha_s$ and a normalized duration parameter $\Delta_d$ as $t_s = \left\lfloor \alpha_s T \right\rfloor$ and $t_e = \min\left(T,\left\lfloor (\alpha_s+\Delta_d)T \right\rfloor\right)$.

The drift may appear early, centrally, or late in the trajectory, and the duration may vary from a localized event to a sustained non-stationary interval based on values of $\alpha_s$ and $\Delta_d$. 

A robust reference scale is computed from the AWGN-corrupted no-drift baseline $y_t^{(i,0)}$ before drift is injected:
\begin{equation}
    \hat{\sigma}_{\mathrm{ref}}
    = 1.4826\cdot
    \operatorname{median}\left(
    \left|y_t^{(i,0)}-\operatorname{median}\left(y_t^{(i,0)}\right)\right|
    \right).
    \label{eq:robust_sigma}
\end{equation}
This median-absolute-deviation-based (MAD) estimate is less sensitive to outliers than the conventional sample standard deviation. It provides a common scale for defining drift strengths across different SNRs and reference trajectories.

The drift types utilised in this study are discussed below:

\subsubsection{Mean-Shift Drift}

Mean-shift drift represents a sustained displacement in the local average level of the observed process. In wireless network traffic and radio access network telemetry, such drift may arise from persistent load changes, baseline bias, long-duration scheduling changes, or slowly varying operating conditions. The mean-shift drift is synthesized by adding a constant offset over $\mathcal{T}_d$: $y_t^{\mathrm{mean}} = y_t^{(i,0)} + c$ for $t \in \mathcal{T}_d$ and $y_t^{\mathrm{mean}} = y_t^{(i,0)}$ for $t \notin \mathcal{T}_d$, with $c = s_{\mu}\kappa_{\mu}\hat{\sigma}_{\mathrm{ref}}$.
Here, $\kappa_{\mu}>0$ is the mean-shift strength factor and $s_{\mu}\in\{-1,+1\}$ controls the direction of the baseline displacement. Positive values simulate upward traffic-level shifts, whereas negative values simulate sustained downward level shifts. Small values of $\kappa_\mu$ generate incipient or weak mean drift, while large values generate strong first-order departures. 

\subsubsection{Variance-Change Drift}

Variance-change drift models local heteroscedasticity, where the spread or power of the corruption process changes while the local mean is preserved. In wireless communication and network traffic monitoring, this may correspond to bursty interference, unstable channel conditions, congestion-induced traffic variability, or local increases in background fluctuations.

The drift is synthesized by adding a zero-mean Gaussian perturbation within $\mathcal{T}_d$: $y_t^{\mathrm{var}} = y_t^{(i,0)} + \tilde{e}_t$ for $t\in\mathcal{T}_d$ and $y_t^{\mathrm{var}} = y_t^{(i,0)}$ for $t\notin\mathcal{T}_d$, where the uncentred perturbation is \(e_t \sim \mathcal{N}(0,\sigma_v^2)\) with local perturbation scale \(\sigma_v = \kappa_{\sigma}\hat{\sigma}_{\mathrm{ref}}\). To avoid inducing an unintended mean shift, the perturbation is sample-mean centred over the drift segment as $\tilde{e}_t = e_t - \tfrac{1}{|\mathcal{T}_d|}\sum_{s\in\mathcal{T}_d} e_s$.
The variance-drift strength range is $\kappa_{\sigma}\in[1.2,6.0]$.
This interval covers weak, moderate, and severe local variance inflation. 

\subsubsection{Structural-Divergence Drift}

This drift alters the empirical distributional shape of the current segment relative to the baseline while preserving its first two moments. This class of drift is relevant when network telemetry undergoes shape changes due to quantization, compression, traffic-composition changes, scheduler behaviour, or other protocol-level regime transitions. Such shifts may not be fully captured by mean or variance statistics.

Let the drift-segment mean and standard deviation of the no-drift baseline be $\mu_d = \tfrac{1}{|\mathcal{T}_d|}\sum_{t\in\mathcal{T}_d} y_t^{(i,0)}$ and $\sigma_d = [\tfrac{1}{|\mathcal{T}_d|}\sum_{t\in\mathcal{T}_d}(y_t^{(i,0)}-\mu_d)^2]^{1/2}$. The segment is first standardized, with $\epsilon_{\mathrm{num}}>0$ a small constant for numerical stability, as $z_t = (y_t^{(i,0)}-\mu_d)/(\sigma_d+\epsilon_{\mathrm{num}})$ for $t\in\mathcal{T}_d$. It is then quantized to $n_{\mathrm{levels}}$ uniformly spaced reconstruction levels, $q_t = \tfrac{1}{n_{\mathrm{levels}}/2}\operatorname{round}(\tfrac{n_{\mathrm{levels}}}{2}z_t)$. The post-quantization mean and standard deviation are $\bar{q}=\tfrac{1}{|\mathcal{T}_d|}\sum_{t\in\mathcal{T}_d}q_t$ and $\hat{\sigma}_q = [\tfrac{1}{|\mathcal{T}_d|}\sum_{t\in\mathcal{T}_d}(q_t-\bar{q})^2]^{1/2}$. Since quantization can slightly alter the segment moments, the quantized samples are renormalized as $\tilde{y}_t = \tfrac{q_t-\bar{q}}{\hat{\sigma}_q+\epsilon_{\mathrm{num}}}\sigma_d+\mu_d$ for $t\in\mathcal{T}_d$. The final structural-divergence signal is $y_t^{\mathrm{str}}=\tilde{y}_t$ for $t\in\mathcal{T}_d$ and $y_t^{\mathrm{str}}=y_t^{(i,0)}$ for $t\notin\mathcal{T}_d$,
with $n_{\mathrm{levels}}\in\{3,5,7,9,12,16\}$ affecting the drift strength. 
The intended statistical effect is $p_t \neq p_{\mathrm{ref}}$ while preserving $\mu_t \approx \mu_{\mathrm{ref}}, \qquad \sigma_t^2 \approx \sigma_{\mathrm{ref}}^2$. 

This construction converts a smooth empirical distribution into a staircase-like distribution while restoring the original local mean and variance. Smaller values of $n_{\mathrm{levels}}$ create stronger distributional deformation because the segment is forced onto fewer reconstruction levels. Larger values create milder structural drift.

\subsubsection{Tail-Amplification Drift}

Tail-amplification drift models the emergence of rare but high-amplitude deviations. In wireless communication and network monitoring, this may correspond to impulsive interference, queue bursts, packet-load spikes, measurement glitches, or heavy-tailed noise. This drift is highly relevant for wavelet denoising because sparse impulses often appear as large localized high-frequency wavelet coefficients.
It is synthesized by superimposing a sparse, bipolar, large-amplitude impulse field on the segment $\mathcal{T}$. 

The sparse impulse field is defined over $\mathcal{T}_d$ as $i_t = s_k A_k$ for $t=t_k$ ($k=1,2,\ldots,K$) and $i_t=0$ otherwise, where the impulse locations $\{t_k\}_{k=1}^{K}$ are sampled uniformly without replacement from $\mathcal{T}_d$. The number of impulses is $K=\max(4,\lfloor \rho_{\mathrm{imp}}|\mathcal{T}_d|\rfloor)$, where $\rho_{\mathrm{imp}}$ is the impulse density, the signs are $s_k\in\{-1,+1\}$, and the impulse amplitudes are sampled as $A_k \sim \operatorname{Uniform}(a_{\min}\hat{\sigma}_{\mathrm{ref}},a_{\max}\hat{\sigma}_{\mathrm{ref}})$. To prevent a net mean displacement, the impulse field is centred, $\tilde{i}_t = i_t - \tfrac{1}{|\mathcal{T}_d|}\sum_{s\in\mathcal{T}_d}i_s$. The tail-amplified signal is then $y_t^{\mathrm{tail}}=y_t^{(i,0)}+\tilde{i}_t$ for $t\in\mathcal{T}_d$ and $y_t^{\mathrm{tail}}=y_t^{(i,0)}$ for $t\notin\mathcal{T}_d$.
The amplitude-factor pairs used here are $(a_{\min},a_{\max})\in\{(4,6),(6,8),(8,10),(10,15)\}$. 
Table~\ref{tab:drift_params} summarizes the parameter ranges adopted for final dataset construction. These ranges are designed to span weak, moderate, and severe drift conditions across multiple SNR regimes and drift locations.

\begin{table}[htbp]
\centering
\caption{Parameter ranges used for drift-synthesis dataset generation.}
\label{tab:drift_params}
\begin{tabular}{lll}
\toprule
\textbf{Component} & \textbf{Symbol} & \textbf{Range / Setting} \\
\midrule
Reference trajectory & $x_t^{(i)}$ & All NST reference trajectories \\
Signal length & $T$ & $701$ \\
AWGN SNR & $\mathrm{SNR}_{\mathrm{dB}}$ & $\{-10,-5,0,5,10,15,20,30\}$ dB \\
Drift start fraction & $\alpha_s$ & $[0.10,0.75]$ \\
Drift duration fraction & $\Delta_d$ & $[0.10,0.35]$ \\
Mean-shift strength & $\kappa_{\mu}$ & $[0.5,3.0]$ \\
Mean-shift sign & $s_{\mu}$ & $\{-1,+1\}$ \\
Variance strength & $\kappa_{\sigma}$ & $[1.2,6.0]$ \\
Quantization levels & $n_{\mathrm{levels}}$ & $\{3,5,7,9,12,16\}$ \\
Impulse density & $\rho_{\mathrm{imp}}$ & $\{0.01,0.03,0.05,0.08,0.10,0.15\}$ \\
Stability constant & $\epsilon_{\mathrm{num}}$ & $10^{-12}$ \\
\bottomrule
\end{tabular}
\end{table}

\subsection{Downstream Tasks: Transient Anomaly Detection and Capacity Estimation}
\label{subsec:tasks}
The denoiser is optimised for two fixed downstream operators applied to the
denoised stream $\hat{x}$. Both reference the \emph{clean} load $x$ for ground
truth, so neither collapses to per-sample reconstruction and neither overlaps the
drift gate.

\emph{Anomaly detection.} A genuine anomaly is a localised, multi-scale
\emph{transient} burst superimposed on the load. Let
$\mathcal{B}\subset\{1,\dots,T\}$ be the burst support in the clean signal; the
ground-truth label is $\ell_t=\mathbb{1}\{t\in\mathcal{B}\}$.
The detector operates on the denoised stream through a residual transient-energy
score against a robust local baseline $\mathrm{med}_w(\cdot)$ (a width-$w$ running
median)~\cite{pacheco2024wavelet},
\begin{equation}
    \zeta_t(\hat{x}) = \big|\,\hat{x}_t - \mathrm{med}_w(\hat{x})_t\,\big|,
    \qquad \hat{\ell}_t(\tau)=\mathbb{1}\{\zeta_t(\hat{x})>\tau\}.
    \label{eq:det_score}
\end{equation}
Detection quality is the area under the ROC curve ($\mathrm{AUC}$), obtained by sweeping $\tau$, together with the $F_1$ score (the harmonic mean of precision and recall) at a fixed operating point. A genuine burst produces a large value of \eqref{eq:det_score}; a
low-pass-smoothed stream does not. The task therefore favours denoisers that
preserve transients while removing noise, which neither a fixed shrinkage rule nor
a moving-average filter achieves across drift regimes. The label
$\boldsymbol{\ell}$ depends on $x$, while the Stage-1 gate depends on the
corruption $n_t+d_t$.

\emph{Capacity estimation.} The operational required capacity is the upper-tail statistic $C_{95}(x)=\mathrm{Pctl}_{95}(x)$, where $\mathrm{Pctl}_{p}(\cdot)$ denotes the empirical $p$th percentile of its argument. The estimate is $\hat{C}_{95}=\mathrm{Pctl}_{95}(\hat{x})$, with error
\begin{equation}
    \mathcal{E}_{\mathrm{cap}} = \big|\hat{C}_{95}-C_{95}\big|,
    \qquad \mathrm{bias} = \hat{C}_{95}-C_{95}.
    \label{eq:cap_err}
\end{equation}
Because $C_{95}$ is a tail statistic, additive noise inflates $\hat{C}_{95}$
(over-provisioning) while over-smoothing deflates it (under-provisioning); a good denoiser must preserve the upper tail without amplifying noise.
The denoising policy is optimised to maximise detection accuracy and minimise capacity error, not per-sample fidelity to the clean window.

\subsection{Framework Overview}
\label{subsec:overview}
\begin{figure}[!t]
  \centering
  \includegraphics[width=\linewidth]{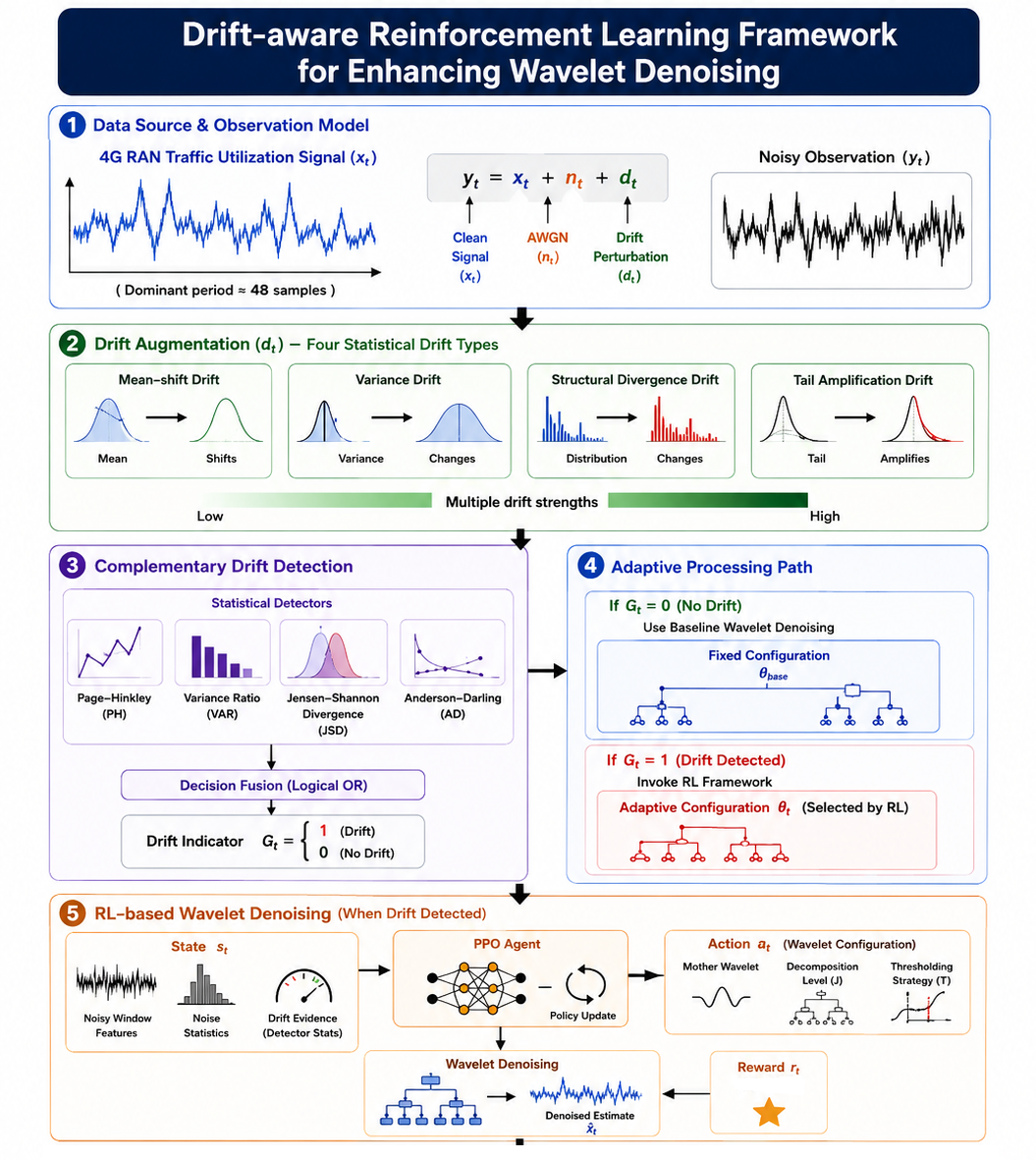}
  \caption{Proposed drift-aware framework: a drift-detection gate routes each window to either a fixed baseline denoiser or a PPO-selected adaptive denoiser; the denoised stream feeds fixed downstream operators (a transient-anomaly detector and a capacity estimator) whose task utility defines the training reward.}
  \label{fig:overview}
\end{figure}

The framework, depicted in Fig.~\ref{fig:overview}, comprises a drift-detection gate followed by an RL-selected adaptive denoiser feeding fixed downstream operators (a transient-anomaly detector and a capacity estimator). A streaming window is processed as follows. First, the noisy telemetry window $\mathbf{y}_m$, which may contain drift in its statistics, is acquired, and drift detection is performed on it to obtain the fused indicator $G_m$. If $G_m=0$ (no drift), the window is denoised with the fixed baseline configuration $\theta_{\mathrm{base}}$ (a VisuShrink setting: Haar wavelet, decomposition depth $J=2$, soft universal thresholding); if $G_m=1$ (drift), the PPO policy is invoked to select an adaptive configuration $\theta_m^{\mathrm{RL}}$. Wavelet denoising is then applied to obtain $\hat{\mathbf{x}}_m$, which is appended to the running denoised history, and the fixed downstream operators are applied to the denoised stream to score the transient-anomaly detector \eqref{eq:det_score} and update the capacity estimate. During training, the reward is the gain in detection accuracy and the reduction in capacity error \eqref{eq:reward_task}, after which the policy/value functions are updated. Finally, the detection flags and the capacity estimate $\hat{C}_{95}$ are emitted to the monitoring system, and processing advances to $\mathbf{y}_{m+1}$.

\section{Drift Detection}
\label{sec:drift_detection}

Prior to adaptive wavelet denoising, the observed signal is examined for statistical drift on a window basis. Let the generated observed trajectory be $y_t = x_t+n_t+d_t$ as per equation ~\ref{eq:obs_model}. 
The full trajectory is segmented into overlapping windows of length \(W\) and hop size \(H\). The \(m\)-th observed window is denoted by $\mathbf{y}_m = [y_{a_m}, y_{a_m+1}, \ldots, y_{a_m+W-1}]^{\top}$, with $a_m = 1+(m-1)H$, and \(m=1,2,\ldots,M\) indexes the window-level decision step. The corresponding reference or baseline window denoted by $\mathbf{y}_{\mathrm{ref},m}$, is obtained from the no-drift baseline 
\(y_t^{(i,0)}\) of the same trajectory and SNR, or from a calibration/rolling baseline in deployment. 

The drift-detection layer has two objectives. First, it determines whether the current window \(\mathbf{y}_m\) departs from the reference statistical regime. Second, it produces detector statistics and binary detector flags that are later embedded into the reinforcement-learning state. This is necessary because the RL policy should not only know whether drift is present, but also whether the departure is primarily mean-based, variance-based, distributional, or tail-sensitive. The framework therefore applies four complementary detectors, $\mathcal{D}_{\mathrm{det}} = \{\mathrm{PH}, \mathrm{VAR}, \mathrm{JSD}, \mathrm{AD}\}$~\cite{jayakumar2026graph}. These detectors correspond respectively to Page-Hinkley, variance-ratio detector, Jensen-Shannon Divergence Detector, and Anderson-Darling Detector.

\subsection{Page-Hinkley Detector for Mean-Shift Drift}

The Page-Hinkley detector is used to detect persistent changes in the first-order moment~\cite{page1954continuous}. Because the mean-shift sign is sampled  \(s_{\mu}\in\{-1,+1\}\), a two-sided statistic is used. 
For the \(m\)-th window, define the scalar sequence supplied to the detector as the local window mean $u_m = \tfrac{1}{W}\sum_{r=0}^{W-1} y_{a_m+r}$. The running mean of this scalar sequence is $\bar{u}_m = \tfrac{1}{m}\sum_{j=1}^{m} u_j$.

Two cumulative deviation processes are maintained, with tolerance \(\delta_{\mathrm{PH}}>0\) and optional forgetting factor \(\alpha_{\mathrm{PH}}\in(0,1]\). Setting \(\alpha_{\mathrm{PH}}=1\) recovers the classical cumulative Page-Hinkley statistic,whereas \(\alpha_{\mathrm{PH}} <1\) discounts older evidence and gives greater weight to recent deviations. 
 
The upward and downward cumulative deviations are $S^{\mathrm{PH},+}_m = \alpha_{\mathrm{PH}} S^{\mathrm{PH},+}_{m-1} + (u_m-\bar{u}_m-\delta_{\mathrm{PH}})$ and $S^{\mathrm{PH},-}_m = \alpha_{\mathrm{PH}} S^{\mathrm{PH},-}_{m-1} + (\bar{u}_m-u_m-\delta_{\mathrm{PH}})$, with corresponding running minima $S^{\mathrm{PH},+}_{m,\min} = \min_{1\leq j\leq m} S^{\mathrm{PH},+}_j$ and $S^{\mathrm{PH},-}_{m,\min} = \min_{1\leq j\leq m} S^{\mathrm{PH},-}_j$. The one-sided and bidirectional PH scores are $\Pi^{\mathrm{PH},+}_m = S^{\mathrm{PH},+}_m - S^{\mathrm{PH},+}_{m,\min}$, $\Pi^{\mathrm{PH},-}_m = S^{\mathrm{PH},-}_m - S^{\mathrm{PH},-}_{m,\min}$, and $\Pi^{\mathrm{PH}}_m = \max(\Pi^{\mathrm{PH},+}_m, \Pi^{\mathrm{PH},-}_m)$.

The binary PH detector flag is $b^{\mathrm{PH}}_m = 1$ if $\Pi^{\mathrm{PH}}_m > h_{\mathrm{PH}}$ and $b^{\mathrm{PH}}_m = 0$ otherwise, where \(h_{\mathrm{PH}}\) is the calibrated Page-Hinkley threshold. 

\subsection{Variance-Ratio Detector for Variance-Change Drift}

Variance-change drift is detected by comparing the second-order statistics of the current window with those of the reference regime. With
$\sigma_m^2 = \operatorname{Var}(\mathbf{y}_m)$,
and
    $\sigma_{\mathrm{ref},m}^2
    =
    \operatorname{Var}(\mathbf{y}_{\mathrm{ref},m})$, 
the variance-ratio statistic is $R_m^{\mathrm{VAR}} = \sigma_m^2/(\sigma_{\mathrm{ref},m}^2+\epsilon_{\mathrm{num}})$. Under no variance drift, \(R_m^{\mathrm{VAR}} \approx 1\). The variance detector flag is $b_m^{\mathrm{VAR}} = 1$ if $|R_m^{\mathrm{VAR}}-1|>h_{\mathrm{VAR}}$ and $b_m^{\mathrm{VAR}} = 0$ otherwise.

\subsection{Jensen-Shannon Divergence Detector for Structural-Divergence Drift}

Let $p_{\mathrm{ref},m}$ and $p_m$ be the normalized empirical distributions of \(\mathbf{y}_{\mathrm{ref},m}\) and \(\mathbf{y}_m\) on common bins. The midpoint distribution is $q_m^{\mathrm{JS}} = \tfrac{1}{2}(p_{\mathrm{ref},m}+p_m)$. The Jensen-Shannon divergence~\cite{lin1991divergence} is $D_m^{\mathrm{JSD}} = \tfrac{1}{2}D_{\mathrm{KL}}(p_{\mathrm{ref},m}\|q_m^{\mathrm{JS}}) + \tfrac{1}{2}D_{\mathrm{KL}}(p_m\|q_m^{\mathrm{JS}})$, where the Kullback--Leibler (KL) divergence is $D_{\mathrm{KL}}(p\|q) = \sum_{\ell} p(\ell)\log\frac{p(\ell)+\epsilon_{\mathrm{num}}}{q(\ell)+\epsilon_{\mathrm{num}}}$. The stabilising constant is added to both numerator and denominator so that \(D_{\mathrm{KL}}(P \parallel p) = 0\) is preserved. The structural-divergence flag is $b_m^{\mathrm{JSD}} = 1$ if $D_m^{\mathrm{JSD}}>h_{\mathrm{JSD}}$ and $b_m^{\mathrm{JSD}} = 0$ otherwise.

\subsection{Anderson-Darling Detector for Tail-Amplification Drift}

Let \(F_{\mathrm{ref},m}(u)\) be the empirical CDF estimated from the reference window \(\mathbf{y}_{\mathrm{ref},m}\) and let $z_{m,(1)} \leq z_{m,(2)} \leq \cdots \leq z_{m,(W)}$ be the ordered samples of the current window \(\mathbf{y}_m\). The two-sample form of the Anderson-Darling statistic~\cite{anderson1952asymptotic} is
\begin{equation}
\begin{split}
A_m^{\mathrm{AD}} = {}& -W - \frac{1}{W}\sum_{r=1}^{W}(2r-1) \Big[\log F_{\mathrm{ref},m}\big(z_{m,(r)}\big) \\
&\qquad {}+ \log\!\big(1-F_{\mathrm{ref},m}\big(z_{m,(W+1-r)}\big)\big)\Big].
\end{split}
\end{equation}
For numerical stability, the CDF values are clipped to the interval \([\epsilon_{\mathrm{num}}, 1-\epsilon_{\mathrm{num}}]\). The tail-amplification flag raises an alarm when $A_m^{\mathrm{AD}} > h_{\mathrm{AD}}$, i.e. $b_m^{\mathrm{AD}} = 1$ if $A_m^{\mathrm{AD}}>h_{\mathrm{AD}}$ and $b_m^{\mathrm{AD}} = 0$ otherwise.

\subsubsection{Threshold Calibration}
\label{subsubsec:threshold_calibration}

The thresholds \(h_{\mathrm{PH}}, h_{\mathrm{VAR}}, h_{\mathrm{JSD}}, h_{\mathrm{AD}}\) are calibrated using only no-drift calibration trajectories $y_t^{(i,0)}=x_t^{(i)}+n_t^{(i)}$ (with $d_t=0$), disjoint from training, validation and test. Let \(Z^d\) denote the corresponding detector statistic at window \(m\), with $Z_m^{\mathrm{PH}} = \Pi_m^{\mathrm{PH}}$, $Z_m^{\mathrm{VAR}} = |R_m^{\mathrm{VAR}} - 1|$, $Z_m^{\mathrm{JSD}} = D_m^{\mathrm{JSD}}$, and $Z_m^{\mathrm{AD}} = A_m^{\mathrm{AD}}$, so that each threshold \(h_d\) is calibrated on the same statistic its flag tests against. For the \(k\)-th calibration trajectory, define $M_k^d = \max_m Z_m^d$. The detector threshold is then selected as $h_d = \gamma_d\, Q_q(\{M_k^d\}_{k=1}^{K_{\mathrm{cal}}})$, where \(Q_q(\cdot)\) is the \(q\)-th empirical percentile, \(\gamma_d\geq1\) is a safety margin, and \(K_{\mathrm{cal}}\) is the number of no-drift calibration trajectories. 

\subsubsection{Fused Detector Evidence}
\label{subsubsec:fused_detector_evidence}

Each detector produces one binary flag and one continuous statistic. The binary detector vector is $\boldsymbol{\phi}_m^{(\mathrm{det,bin})} = [b_m^{\mathrm{PH}}, b_m^{\mathrm{VAR}}, b_m^{\mathrm{JSD}}, b_m^{\mathrm{AD}}]$ and the continuous detector-statistic vector is $\boldsymbol{\phi}_m^{(\mathrm{det,stat})} = [\Pi_m^{\mathrm{PH}}, R_m^{\mathrm{VAR}}, D_m^{\mathrm{JSD}}, A_m^{\mathrm{AD}}]$. The window-level fused drift indicator is constructed using logical OR, $G_m = b_m^{\mathrm{PH}} \vee b_m^{\mathrm{VAR}} \vee b_m^{\mathrm{JSD}} \vee b_m^{\mathrm{AD}}$, so that $G_m=1$ when at least one detector flags drift in window $m$ and $G_m=0$ when no detector flags drift. The complete detector-evidence vector supplied to the RL state is
\begin{equation}
    \boldsymbol{\phi}_m^{(\mathrm{det})}
    =
    [
    b_m^{\mathrm{PH}},
    b_m^{\mathrm{VAR}},
    b_m^{\mathrm{JSD}},
    b_m^{\mathrm{AD}},
    G_m,
    \Pi_m^{\mathrm{PH}},
    R_m^{\mathrm{VAR}},
    D_m^{\mathrm{JSD}},
    A_m^{\mathrm{AD}}
    \label{eq:detector_vector}
    ].
\end{equation}

At the trajectory level, the signal is treated as drift-affected if any window fires, $G_{\mathrm{sig}} = \bigvee_{m=1}^{M} G_m$. The decision index \(m\) corresponds
to the \(m\)-th overlapping window of the observed trajectory, \(\mathbf{y}_m\) denotes
the current observed window.
If \(G_{\mathrm{sig}}=0\), the trajectory is denoised with the baseline configuration \(\theta_{\mathrm{base}}\); if \(G_{\mathrm{sig}}=1\), RL policy is activated. The precise per-window versus per-trajectory routing is specified in Section IV.

\section{Drift-Aware Adaptive Wavelet Denoising via Reinforcement Learning}
\label{sec:rl}

\subsection{Reinforcement-Learning Formulation}

The standard Markov Decision Process (MDP) tuple is formulated as $\mathcal{M} = (\mathcal{S},\mathcal{A},\mathcal{P},\mathcal{R},\gamma)$, where \(\mathcal{S}\) is the state space, \(\mathcal{A}\) is the action space, \(\mathcal{P}(s_{m+1}\mid s_m,a_m)\) is the transition kernel, \(\mathcal{R}(s_m,a_m,s_{m+1})\) is the reward function, and \(\gamma\in(0,1)\) is the discount factor~\cite{puterman2014markov}. The RL formulation used in the per-window decision studied here is its one-step (contextual-bandit)~\cite{lattimore2020bandit} special case $\mathcal{M} = (\mathcal{S}, \mathcal{A}, R)$.

The policy is invoked only where drift is present. With the trajectory-level indicator \(G_{\mathrm{sig}}\) of Section III, the configuration applied at window \(m\) is $\theta_m = \theta_{\mathrm{base}}$ if $G_{\mathrm{sig}}=0$ and $\theta_m = \theta_m^{\mathrm{RL}}$ if $G_{\mathrm{sig}}=1$, where \(\theta_{\mathrm{base}}\) is the fixed baseline (VisuShrink) configuration and \(\theta_{m}^{\mathrm{RL}}\) is selected by the policy. The agent learns \(\theta_{\mathrm{sig}}^{\mathrm{RL}}\) so that denoising remains effective under non-stationary noise and drift. 

\subsection{State Representation}
\label{subsubsec:state_representation}

The state at decision step \(m\) is constructed by combining signal-level descriptors, change-sensitive descriptors, and detector-derived drift evidence: $s_m = [\boldsymbol{\phi}_m^{(\mathrm{sig})}, \boldsymbol{\phi}_m^{(\mathrm{chg})}, \boldsymbol{\phi}_m^{(\mathrm{det})}]$, where \(\boldsymbol{\phi}_m^{(\mathrm{sig})}\) contains features extracted directly from the observed window \(\mathbf{y}_m\), \(\boldsymbol{\phi}_m^{(\mathrm{chg})}\) contains temporal or historical change descriptors, and \(\boldsymbol{\phi}_m^{(\mathrm{det})}\) contains detector flags and detector statistics.

A representative signal-feature vector may include local amplitude and shape descriptors of the observed window, $\boldsymbol{\phi}_m^{(\mathrm{sig})} = [\mu_m, \sigma_m, \operatorname{MAD}_m, y_m^{\min}, y_m^{\max}, r_m,$ $\|\mathbf{y}_m\|_2, \operatorname{skew}(\mathbf{y}_m), \operatorname{kurt}(\mathbf{y}_m), \operatorname{PAR}_m]$. The local mean, variance and median absolute deviation are $\mu_m = \tfrac{1}{W}\sum_{r=0}^{W-1} y_{a_m+r}$, $\sigma_m^2 = \operatorname{Var}(\mathbf{y}_m)$, and $\operatorname{MAD}_m = \operatorname{median}(|\mathbf{y}_m - \operatorname{median}(\mathbf{y}_m)|)$. The dynamic range and peak-to-average-ratio features are $r_m = y_m^{\max} - y_m^{\min}$ and $\operatorname{PAR}_m = \max_r |y_{a_m+r}|/(\sqrt{\tfrac{1}{W}\sum_{r=0}^{W-1} y_{a_m+r}^{2}} + \epsilon_{\mathrm{num}})$.

The change-sensitive feature vector \(\boldsymbol{\phi}_m^{(\mathrm{chg})}\) summarizes short-term departures from recent operating history, $\boldsymbol{\phi}_m^{(\mathrm{chg})} = [\Delta \mu_m, \rho_{\sigma,m}, \bar{R}^{\mathrm{VAR}}_m, \bar{\Pi}^{\mathrm{PH}}_m, \bar{D}^{\mathrm{JSD}}_m]$, with $\Delta \mu_m = (\mu_m-\bar{\mu}_{m-1})/(|\bar{\mu}_{m-1}|+\epsilon_{\mathrm{num}})$ and $\rho_{\sigma,m} = \sigma_m/(\bar{\sigma}_{m-1}+\epsilon_{\mathrm{num}})$, where \(\bar{\mu}_{m-1}\) and \(\bar{\sigma}_{m-1}\) are running estimates computed from previous windows.
The detector-evidence vector inherited from the drift-detection stage \eqref{eq:detector_vector} exposes both discrete detector decisions and continuous detector statistics to the RL agent. Consequently, the policy can condition its wavelet-denoising action on drift presence, drift type, and drift intensity.

Because the components of \(s_m\) may have different physical units and numerical ranges, feature normalization is applied before the state is passed to the policy. Let \(z_{m,k}\) denote the \(k\)-th raw state feature at decision step \(m\). The z-score normalized feature and the normalised RL state are $\tilde{z}_{m,k} = (z_{m,k}-\mu_k)/(\sigma_k+\epsilon_{\mathrm{num}})$ and $\tilde{s}_m = \mathcal{N}(s_m)$, where \(\mathcal{N}(\cdot)\) denotes the normalization operator. 

\subsection{Action Space}
\label{subsubsec:action_space}

The RL action selects the adaptive wavelet-denoising configuration for the current window. The action at decision step \(m\) is $a_m = \theta_m^{\mathrm{RL}} = (\psi_m, J_m, \kappa_m, \tau_m, \alpha_m, \rho_m)$, drawn from the mixed discrete-continuous action space $\mathcal{A} = \mathcal{A}_{\psi} \times \mathcal{A}_{J} \times \mathcal{A}_{\kappa} \times \mathcal{A}_{\tau} \times \mathcal{A}_{\alpha} \times \mathcal{A}_{\rho}$, where the categorical components are the selected mother wavelet \(\psi_m\), the decomposition depth \(J_m\), the thresholding function \(\kappa_m\) and the threshold-selection strategy \(\tau_m\). The continuous components are a threshold-gain parameter \(\alpha_m\) and a per-scale shaping parameter \(\rho_m\), satisfying \(\alpha_m \in [\alpha_{\min}, \alpha_{\max}]\) and \(\rho_m \in [\rho_{\min}, \rho_{\max}]\).

The selected action is decoded into a wavelet-denoising operator that maps the observed window \(\mathbf{y}_m\) to the denoised estimate
\begin{equation}
\hat{\mathbf{x}}_m = \mathcal{W}^{-1}_{\psi_m,J_m}\big[\mathcal{T}_{\kappa_m,\boldsymbol{\lambda}_m}(\mathcal{W}_{\psi_m,J_m}(\mathbf{y}_m))\big],
\end{equation}
where \(\mathcal{W}_{\psi_m,J_m}(\cdot)\) and \(\mathcal{W}^{-1}_{\psi_m,J_m}(\cdot)\) are the forward and inverse wavelet transforms and \(\mathcal{T}_{\kappa_m,\boldsymbol{\lambda}_m}(\cdot)\) applies the selected thresholding rule with scale-dependent thresholds \(\boldsymbol{\lambda}_m\).

At scale \(j\), the threshold may be written as $\lambda_{m,j} = \Lambda_{\tau_m}(\hat{\sigma}_{m,j}, N_{m,j}, \alpha_m, \rho_m)$, where \(\hat{\sigma}_{m,j}\) is the estimated noise scale at wavelet subband \(j\), \(N_{m,j}\) is the number of coefficients at that scale, and \(\Lambda_{\tau_m}(\cdot)\) denotes the threshold-selection mapping associated with strategy \(\tau_m\).
For the universal-threshold case, the base threshold is $\lambda_{m,j} = \alpha_m \hat{\sigma}_{m,j} \sqrt{2\log N_{m,j}}$. A scale-shaping parameter \(\rho_m\) may be introduced as $\lambda_{m,j} = \alpha_m \hat{\sigma}_{m,j} \sqrt{2\log N_{m,j}}\,[1+\rho_m \eta_j]$, where \(\eta_j\in[0,1]\) is a normalized scale index. Thus, \(\rho_m\) lets the policy tilt threshold aggressiveness from fine to coarse scales. The thresholding function \(\kappa_m\) determines how each coefficient is modified: the soft and hard thresholding operators are $T_{\lambda}^{\mathrm{soft}}(w) = \operatorname{sign}(w)\max(|w|-\lambda,0)$ and $T_{\lambda}^{\mathrm{hard}}(w) = w\,\mathbf{1}\{|w| \geq \lambda\}$. Thus, \(\tau_m\) controls how the threshold is computed, while \(\kappa_m\) controls how the coefficient is transformed. The best wavelet basis depends on the noise regime  differing for Gaussian and impulsive corruption~\cite{dey2022wavelet}.

\subsection{Fixed Downstream Operators}
\label{subsec:frozen_forecaster}
The reward defined below depends on the two downstream operators of
Section~\ref{subsec:tasks}. These operators are kept \emph{fixed} (not learned jointly with the
policy), thereby maintaining the reward as stationary. If either operator were optimised under the agent, the reward would be non-stationary through both data drift and a moving operator,
which is detrimental to on-policy methods such as PPO. Both these operators are parameter-light and require no training: the anomaly detector is the residual transient-energy score \eqref{eq:det_score} at a fixed threshold and the capacity
estimator is an upper-tail percentile. The detector thresholds and the percentile levels are calibrated once on the clean training references and then held fixed across policy learning and evaluation. Thus, the gate determines whether the policy is invoked, whereas the fixed downstream operators evaluate the recovered load. If a learned detector or estimator is used instead, it is pretrained on clean references and frozen before policy learning.

\subsection{Reward Design}
\label{subsubsec:reward_design}
The agent's \emph{primary} reward combines a detection term and a capacity term evaluated by
the fixed operators of Section~\ref{subsec:frozen_forecaster}. Let $\hat{\mathbf{x}}_m$ denote the denoised window produced by the selected configuration and $\mathbf{y}_m$ the corresponding noisy window. The detection term is the gain in window-level detection $F_1$ of the denoised stream over the raw observation,
\begin{equation}
r_m^{\mathrm{det}} = F_1(\hat{\mathbf{x}}_m,\boldsymbol{\ell}_m) - F_1(\mathbf{y}_m,\boldsymbol{\ell}_m),
\end{equation}
and the capacity term is the reduction in normalised capacity error,
\begin{equation}
r_m^{\mathrm{cap}} = \frac{|\,\mathrm{Pctl}_{95}(\mathbf{y}_m)-C_{95}^{m}\,| - |\,\mathrm{Pctl}_{95}(\hat{\mathbf{x}}_m)-C_{95}^{m}\,|}{C_{95}^{m}+\epsilon_{\mathrm{num}}},
\end{equation}
where $C_{95}^{m}=\mathrm{Pctl}_{95}(\mathbf{x}_m)$.

The task reward is their convex combination
\begin{equation}
    r_m^{\mathrm{task}} = w_{\mathrm{det}}\, r_m^{\mathrm{det}} + w_{\mathrm{cap}}\, r_m^{\mathrm{cap}},
    \qquad
    w_{\mathrm{det}}+w_{\mathrm{cap}}=1  
    \label{eq:reward_task}
\end{equation}

 with weights $w_{\mathrm{det}},w_{\mathrm{cap}}\ge 0$ trading off the two monitoring objectives (equal weights unless stated). Maximising $r_m^{\mathrm{task}}$ drives
the policy to remove noise and drift artefacts that most damage detection and
capacity estimation, while discouraging over-smoothing that would erase genuine
load excursions or deflate the upper tail.

For comparison and ablation, three fidelity-oriented rewards are retained. With $\mathbf{x}_m$ as the aligned clean window, the SNR-improvement reward is
\begin{equation}
\begin{split}
r_m^{\mathrm{snr}} &= 10\log_{10}\!\left(\frac{\|\mathbf{x}_m\|_2^2}{\|\mathbf{x}_m-\hat{\mathbf{x}}_m\|_2^2+\epsilon_{\mathrm{num}}}\right) \\
&\quad - 10\log_{10}\!\left(\frac{\|\mathbf{x}_m\|_2^2}{\|\mathbf{x}_m-\mathbf{y}_m\|_2^2+\epsilon_{\mathrm{num}}}\right)
\end{split}
\end{equation}
The normalised mean-squared-error (MSE) reward is $r_m^{\mathrm{MSE}} = -\|\mathbf{x}_m-\hat{\mathbf{x}}_m\|_2^2/(\|\mathbf{x}_m\|_2^2+\epsilon_{\mathrm{num}})$;
the power-consistency reward is
$r_m^{\mathrm{pow}} = -\big|\,\|\hat{\mathbf{x}}_m\|_2^2-\|\mathbf{x}_m\|_2^2\,\big|/(\|\mathbf{x}_m\|_2^2+\epsilon_{\mathrm{num}})$. These fidelity rewards optimise reconstruction quantities and  are compared as candidate training objectives in Section~\ref{sec:results}.

\subsection{Learning Objective and Evaluation Metrics}
\label{subsubsec:learning_objective}
The policy $\pi_{\omega}(a_m\mid\tilde{s}_m)$ maps the normalised state to a distribution over mixed wavelet actions and maximises the expected return $J(\pi_\omega) = \mathbb{E}_{s_m \sim \mathcal{D},\, a_m \sim \pi_\omega(\cdot\mid \tilde{s}_m)}[\, r^{\mathrm{task}}_m \,]$.

The \emph{primary} evaluation metrics are the two task metrics on the denoised stream: transient anomaly-detection ROC AUC and $F_1$, and the capacity error \eqref{eq:cap_err}, each broken down by drift type and input SNR. As an \emph{enabling-layer diagnostic}, the denoising SNR improvement $\Delta \mathrm{SNR}_{m} = \mathrm{SNR}_{\mathrm{after},m} - \mathrm{SNR}_{\mathrm{before},m}$ is also reported, which characterises the denoiser independently of the downstream operators. Training reward and evaluation metrics are separated: the agent is trained on $r_m^{\mathrm{task}}$ while AUC, $F_1$, $\mathcal{E}_{\mathrm{cap}}$ and $\Delta\operatorname{SNR}$ are used for reporting.

\subsection{Policy Optimization Using Proximal Policy Optimization}

The adaptive denoising policy is optimized using Proximal Policy Optimization
(PPO)~\cite{schulman2017proximal}. Let \(\pi_{\omega}(a_m|\tilde{s}_m)\) denote the current policy and
\(\pi_{\omega_{\mathrm{old}}}(a_m|\tilde{s}_m)\) denote the policy before the
current update. The PPO probability ratio is $\eta_m(\omega) = \pi_{\omega}(a_m|\tilde{s}_m)/\pi_{\omega_{\mathrm{old}}}(a_m|\tilde{s}_m)$. The clipped PPO objective is
\begin{equation}
\begin{split}
L^{\mathrm{PPO}}(\omega) = \mathbb{E}_m\big[\min\big(&\eta_m(\omega)\hat{A}_m, \\
&\operatorname{clip}(\eta_m(\omega), 1-\epsilon_{\mathrm{clip}}, 1+\epsilon_{\mathrm{clip}})\hat{A}_m\big)\big],
\end{split}
\end{equation}
where \(\hat{A}_m\) is the advantage estimate and \(\epsilon_{\mathrm{clip}}\) is the PPO clipping parameter. The clipping operation restricts the magnitude of policy updates and improves stability under non-stationary drift conditions. The advantage may be estimated as $\hat{A}_m = \hat{Q}(\tilde{s}_m,a_m) - V_{\nu}(\tilde{s}_m)$, where \(V_{\nu}(\tilde{s}_m)\) is the critic's value estimate. 

\subsubsection{Actor-Critic Parameterization}
\label{subsubsec:actor_critic_parameterization}

The PPO agent follows an actor-critic architecture. A shared encoder first maps
the normalized state \(\tilde{s}_m\) into a latent representation \(h_m = f_{\xi}(\tilde{s}_m)\) where \(f_{\xi}(\cdot)\) is a neural feature encoder. The actor then produces
action distributions for the categorical and continuous components of \(a_m\).
The categorical components are modeled as $\psi_m \sim \operatorname{Cat}(\mathbf{p}_{\psi}(\mathbf{h}_m))$, $J_m \sim \operatorname{Cat}(\mathbf{p}_{J}(\mathbf{h}_m))$, $\kappa_m \sim \operatorname{Cat}(\mathbf{p}_{\kappa}(\mathbf{h}_m))$, and $\tau_m \sim \operatorname{Cat}(\mathbf{p}_{\tau}(\mathbf{h}_m))$.

The continuous components are generated using bounded distributions. One practical approach is to sample unconstrained Gaussian variables and map them to bounded intervals using a hyperbolic-tangent transformation: $\alpha_m = \tfrac{\alpha_{\max}-\alpha_{\min}}{2}[\tanh(u_m^{\alpha})+1] + \alpha_{\min}$ and $\rho_m = \tfrac{\rho_{\max}-\rho_{\min}}{2}[\tanh(u_m^{\rho})+1] + \rho_{\min}$, with $u_m^{\alpha} \sim \mathcal{N}(\mu_{\alpha}(\mathbf{h}_m), \sigma_{\alpha}^{2})$ and $u_m^{\rho} \sim \mathcal{N}(\mu_{\rho}(\mathbf{h}_m), \sigma_{\rho}^{2})$.

The critic estimates the value function \(V_{\nu}(\tilde{s}_m) = g_{\nu}(h_m)\)
where \(g_{\nu}(\cdot)\) is the value head. The actor parameter \(\omega\) and
critic parameter \(\nu\) are optimized jointly with the PPO policy loss, value
loss, and entropy regularization. A typical combined training objective is
\begin{equation}
\begin{split}
    \mathcal{L}(\omega,\nu)
    =
    &
    -
    L^{\mathrm{PPO}}(\omega)
    +
    c_v
    \mathbb{E}_m
    \left[
    \left(
    V_{\nu}(\tilde{s}_m)
    -
    \hat{R}_m
    \right)^2
    \right]
    \\
    &
    -
    c_e
    \mathbb{E}_m
    \left[
    \mathcal{H}
    (
    \pi_{\omega}(\cdot|\tilde{s}_m)
    )
    \right],
    \label{L_PPO}
\end{split}
\end{equation}
where \(c_v\) and \(c_e\) are the value-loss and entropy coefficients,
respectively, \(\hat{R}_m\) is the empirical return, and \(\mathcal{H}(\cdot)\)
denotes policy entropy.

\subsection{Training, Validation, and Held-Out Testing Protocol}
\label{subsubsec:training_validation_test_protocol}

The RL training protocol  preserves the separation between detector calibration, policy learning, model selection, and final testing. The dataset is partitioned as $\mathcal{D} = \mathcal{D}_{\mathrm{cal}} \cup \mathcal{D}_{\mathrm{train}} \cup \mathcal{D}_{\mathrm{val}} \cup \mathcal{D}_{\mathrm{test}}$, with detector thresholds calibrated on \(\mathcal{D}_{\mathrm{cal}}\) and then frozen;
policy-gradient updates on \(\mathcal{D}_{\mathrm{train}}\);
hyperparameters, checkpoint selection, reward variant, and state/action design chosen on
\(\mathcal{D}_{\mathrm{val}}\); and a single final evaluation on
\(\mathcal{D}_{\mathrm{test}}\). The fixed downstream operators are calibrated on clean
training references and frozen before any policy update, so no validation or test information leaks through the operators. The held-out test split is used once, under the protocol of
Section~\ref{subsec:forecasting_results}.


\subsection{Algorithmic Summary}
\label{subsubsec:algorithmic_summary}

The framework proceeds in three algorithmic stages. In the first stage, each clean reference trajectory is corrupted with additive white Gaussian noise and one of four drift types (mean shift, variance change, structural divergence, tail amplification). The second stage  integrates four drift detectors (PH,VAR,JSD,AD) into a single decision gate. In the final stage, non-drifted signals are processed using fixed VisuShrink denoising, while drifted signals trigger a PPO agent that configures the wavelet parameters. By rewarding anomaly-detection accuracy and capacity estimation error, the framework learns operationally optimal denoising policies.

\begin{algorithm}[!t]
\scriptsize
\caption{Dataset Generation for Drift-Aware Wavelet Denoising}
\label{alg:dataset_generation}

\Require{
reference trajectories \(\{x_t^{(i)}\}_{i=1}^{N_{\mathrm{ref}}}\), signal length \(T\),
SNR set \(\mathcal{S}_{\mathrm{SNR}}\), drift-type set
\(\mathcal{D}_{\mathrm{drift}}=\{\mathrm{mean},\mathrm{var},\mathrm{str},\mathrm{tail}\}\),
drift-parameter ranges, and random seed set
}

\Ensure{
dataset \(\mathcal{D}\) containing clean signals, no-drift noisy baselines,
drifted signals, drift masks, SNR labels, drift labels, and split labels
}

\For{each reference trajectory \(x_t^{(i)}\)}{

    \For{each \(\mathrm{SNR}_{\mathrm{dB}}\in\mathcal{S}_{\mathrm{SNR}}\)}{

        Compute signal power $P_x^{(i)}$

        Compute AWGN variance $\sigma_n^2$
        
        Generate AWGN \[n_t^{(i)}\sim\mathcal{N}(0,\sigma_n^2).\]
        
        Form the no-drift noisy baseline
        \[
        y_t^{(i,0)}=x_t^{(i)}+n_t^{(i)} .
        \]

        Compute the robust reference scale
        \(\hat{\sigma}_{\mathrm{ref}}\) \eqref{eq:robust_sigma}

        Store the no-drift instance with \(d_t=0\)\;

        \For{each drift type \(d\in\mathcal{D}_{\mathrm{drift}}\)}{

            Sample drift onset fraction \(\alpha_s\) and duration fraction
            \(\Delta_d\)\;

            Compute drift interval
            \[
            t_s=\lfloor \alpha_s T\rfloor,
            \qquad
            t_e=\min\{T,\lfloor(\alpha_s+\Delta_d)T\rfloor\}.
            \]

            Define
            \[
            \mathcal{T}_d=\{t_s,t_s+1,\ldots,t_e\}.
            \]

            \eIf{\(d=\mathrm{mean}\)}{
                Sample \(\kappa_\mu\) and \(s_\mu\in\{-1,+1\}\)\;
                Generate \(y_t^{\mathrm{mean}}\) using the mean-shift equation\;
            }{
            \eIf{\(d=\mathrm{var}\)}{
                Sample \(\kappa_\sigma\)\;
                Generate \(y_t^{\mathrm{var}}\) using zero-mean local variance inflation\;
            }{
            \eIf{\(d=\mathrm{str}\)}{
                Sample \(n_{\mathrm{levels}}\)\;
                Generate \(y_t^{\mathrm{str}}\) using moment-preserving quantization\;
            }{
                Sample \(\rho_{\mathrm{imp}}\) and \((a_{\min},a_{\max})\)\;
                Generate \(y_t^{\mathrm{tail}}\) using sparse centred impulses\;
            }}}

            Store the drifted signal \(y_t\), drift mask, drift type, SNR,
            drift parameters, \((t_s,t_e)\), and seed\;
        }
    }
}

Assign each generated trajectory to calibration, training, validation, or
held-out test split using disjoint reference identifiers and seeds\;

\Return{\(\mathcal{D}\)}\;

\end{algorithm}

\begin{algorithm}[!t]
\scriptsize
\caption{First-Stage Drift Detection}
\label{alg:drift_detection_stage}

\Require{
dataset \(\mathcal{D}\), detector set
\(\mathcal{D}_{\mathrm{det}}=\{\mathrm{PH},\mathrm{VAR},\mathrm{JSD},\mathrm{AD}\}\),
window length \(W\), hop size \(H\), percentile \(q\), safety margin \(\gamma_d\),
and numerical constant \(\epsilon_{\mathrm{num}}\)
}

\Ensure{
window-level detector outputs
\(\{b_m^{\mathrm{PH}},b_m^{\mathrm{VAR}},b_m^{\mathrm{JSD}},b_m^{\mathrm{AD}},G_m\}_{m=1}^{M}\),
signal-level drift trigger \(G_{\mathrm{sig}}\), and drift-detected subset
\(\mathcal{D}_{\mathrm{RL}}\)
}

\textbf{Detector calibration:}\;

Select only no-drift calibration trajectories:
\[
\mathcal{D}_{\mathrm{cal},0}
=
\left\{
y_t^{(i,0)}:
d_t=0,\;
y_t^{(i,0)}\notin
\mathcal{D}_{\mathrm{val}}\cup\mathcal{D}_{\mathrm{test}}
\right\}.
\]

\For{each detector \(d\in\{\mathrm{PH},\mathrm{VAR},\mathrm{JSD},\mathrm{AD}\}\)}{

    \For{each no-drift calibration trajectory \(k\)}{

        Segment the trajectory into overlapping windows
        \[
        \mathbf{y}_m=
        [y_{a_m},y_{a_m+1},\ldots,y_{a_m+W-1}]^{\top},
        \qquad
        a_m=1+(m-1)H .
        \]

        Compute the detector statistic \(Z_m^d\) over all windows\;

        Store the maximum no-drift detector statistic
        \[
        M_k^d=\max_m Z_m^d .
        \]
    }

    Calibrate the detector threshold
    \[
    h_d
    =
    \gamma_d
    Q_q
    \left(
    \{M_k^d\}_{k=1}^{K_{\mathrm{cal}}}
    \right).
    \]
}

Freeze \(\{h_{\mathrm{PH}},h_{\mathrm{VAR}},h_{\mathrm{JSD}},h_{\mathrm{AD}}\}\)

\textbf{Window-wise drift detection:}\;

\For{each non-calibration trajectory \(y_t\in\mathcal{D}\setminus\mathcal{D}_{\mathrm{cal},0}\)}{

    Segment \(y_t\) into windows \(\{\mathbf{y}_m\}_{m=1}^{M}\)\;

    \For{\(m=1\) \KwTo \(M\)}{

        Compute detector statistics \(\{\Pi_m^{\mathrm{PH}}, R_m^{\mathrm{VAR}}, D_m^{\mathrm{JSD}}, A_m^{\mathrm{AD}}\}\).

        Compute detector flags by thresholding each detector statistic (Section~\ref{sec:drift_detection})
        \[
        b_m^{\mathrm{PH}},
        \quad
        b_m^{\mathrm{VAR}},
        \quad
        b_m^{\mathrm{JSD}},
        \quad
        b_m^{\mathrm{AD}} .
        \]

        Fuse the detector flags using logical OR:
        \[
        G_m
        =
        b_m^{\mathrm{PH}}
        \vee
        b_m^{\mathrm{VAR}}
        \vee
        b_m^{\mathrm{JSD}}
        \vee
        b_m^{\mathrm{AD}} .
        \]

        Form the detector-evidence vector
        \[
        \boldsymbol{\phi}_m^{(\mathrm{det})}
        =
        [
        b_m^{\mathrm{PH}},
        b_m^{\mathrm{VAR}},
        b_m^{\mathrm{JSD}},
        b_m^{\mathrm{AD}},
        G_m,
        \Pi_m^{\mathrm{PH}},
        R_m^{\mathrm{VAR}},
        D_m^{\mathrm{JSD}},
        A_m^{\mathrm{AD}}
        ] .
        \]
    }

    Compute the signal-level drift trigger
    \[
    G_{\mathrm{sig}}
    =
    \bigvee_{m=1}^{M}G_m .
    \]

    \eIf{\(G_{\mathrm{sig}}=1\)}{
        Add the trajectory and its window-level detector evidence to
        \(\mathcal{D}_{\mathrm{RL}}\)\;
    }{
        Mark the trajectory for baseline wavelet denoising using
        \(\theta_{\mathrm{base}}\)\;
    }
}

\Return{\(\mathcal{D}_{\mathrm{RL}}\), all detector flags, detector statistics, and \(G_{\mathrm{sig}}\)}\;

\end{algorithm}

\begin{algorithm}[!t]
\scriptsize
\caption{Task-Reward RL Training with Fixed Downstream Operators}
\label{alg:rl_forecast_reward}

\Require{drift-detected training subset $\mathcal{D}_{\mathrm{RL}}$, validation/test subsets, clean references for operator calibration, burst labels \(\ell_m = \mathbf{1}\{t \in \mathcal{B}\}\), baseline $\theta_{\mathrm{base}}$, policy $\pi_{\omega}$, critic $V_{\nu}$, wavelet library $\Psi$, sets $\mathcal{J},\mathcal{K},\mathcal{T}$, action ranges $[\alpha_{\min},\alpha_{\max}],[\rho_{\min},\rho_{\max}]$, normalizer $\mathcal{N}(\cdot)$, reward weights $w_{\mathrm{det}},w_{\mathrm{cap}}$, PPO hyperparameters}
\Ensure{trained policy $\pi_{\omega}$ and held-out detection/capacity scores}

\textbf{Stage~A: operator calibration}\;
Calibrate the detector threshold and capacity level on clean training references; fix them\;

\textbf{Stage~B: policy learning}\;
Initialize PPO actor, critic, rollout buffer, and state normalizer\;
\For{$\mathrm{episode}=1$ \KwTo $N_{\mathrm{episodes}}$}{
  Sample one drift-detected training trajectory from $\mathcal{D}_{\mathrm{RL}}$\;
  Initialise the denoised history buffer\;
  \For{$m=1$ \KwTo $M$}{
    Construct and normalize the state $\tilde{s}_m=\mathcal{N}\big([\boldsymbol{\phi}_m^{(\mathrm{sig})},\boldsymbol{\phi}_m^{(\mathrm{chg})},\boldsymbol{\phi}_m^{(\mathrm{det})}]\big)$\;
    Sample the wavelet action $a_m=(\psi_m,J_m,\kappa_m,\tau_m,\alpha_m,\rho_m)\sim\pi_{\omega}(\cdot\mid\tilde{s}_m)$\;
    Denoise the window $\hat{\mathbf{x}}_m=\mathcal{W}^{-1}_{\psi_m,J_m}\!\big[\mathcal{T}_{\kappa_m,\boldsymbol{\lambda}_m}(\mathcal{W}_{\psi_m,J_m}(\mathbf{y}_m))\big]$ and append to the denoised history\;
    Apply the fixed operators to $\hat{\mathbf{x}}_m$: detection $F_1$ against burst labels and capacity error against $C_{95}^{m}$\;
    Compute the task reward $r_m^{\mathrm{task}}$ from \eqref{eq:reward_task}\;
    Store $(\tilde{s}_m,a_m,r_m^{\mathrm{task}},\mathrm{done})$ in the rollout buffer\;
  }
  \If{rollout buffer is ready}{
    Compute advantages $\hat{A}_m$\;
    Update actor and critic on the combined objective\eqref{L_PPO}:$\mathcal{L}(\omega,\nu)$\;
  }
  Periodically evaluate validation detection AUC and capacity error; save the best checkpoint\;
}

\textbf{Held-out evaluation.}\;
Load the best checkpoint; freeze the normalizer and all operator thresholds\;
\For{each held-out test trajectory}{
  Detect drift and compute $G_{\mathrm{sig}}$\;
  Produce denoised streams for the arms: clean (oracle), raw $\mathbf{y}$, fixed VisuShrink $\theta_{\mathrm{base}}$, SureShrink, BayesShrink, Wiener filter  and policy ($\theta_{\mathrm{base}}$ if $G_{\mathrm{sig}}=0$, else $a_m^{\ast}=\arg\max_a\pi_{\omega}(a\mid\tilde{s}_m)$ with mean continuous actions)\;
  Score each arm with the fixed operators: ROC AUC, $F_1$, and capacity error $|\hat{C}_{95}-C_{95}|$\;
}
\Return{$\pi_{\omega}$ and the per-arm detection/capacity scores}\;
\end{algorithm}

\section{Experiments and Results}
\label{sec:results}

The augmented dataset contains both noisy drifted signals for RL interaction and the ground-truth clean references needed for reward computation and evaluation; augmentation yields $51$ reference trajectories of length $T=701$. Reference trajectories are corrupted by AWGN at eight input SNR levels and modulated by four canonical drifts (mean shift, variance, structural, tail amplification). Held-out testing uses $N_{\mathrm{test}}=288$ instances ($256$ drifted, $32$ no-drift controls) drawn from references disjoint from the training and validation pools. The detector thresholds are calibrated once on the training split's no-drift instances and shared across seeds $\{42, 123, 2025, 7, 11\}$.

\emph{Note on the reported numbers.} All quantitative results in this section are
produced by our evaluation harness, released with the paper. In the present draft
the task numbers are computed on a \emph{synthetic stand-in} signal that
reproduces the seasonal structure of the traffic trace described in
Section~\ref{sec:sysmodel}, because the original trace and the trained PPO policy
were not available at the time of writing. Accordingly, the Proposed arm in
the task comparison is the oracle per-window configuration-search
\emph{upper bound} over the action space (marked $\dagger$) rather than the
learned policy. Running the same harness on the real references and PPO
checkpoint regenerates every figure and table in this section verbatim with the
final numbers; the present results are methodological previews and are not the
final reported performance. The drift-detection rates reported next are from the
authors' original experimental pipeline.

\subsection{Drift Detection Performance}

\begin{figure}[htbp]
  \centering
  \includegraphics[width=0.9\linewidth]{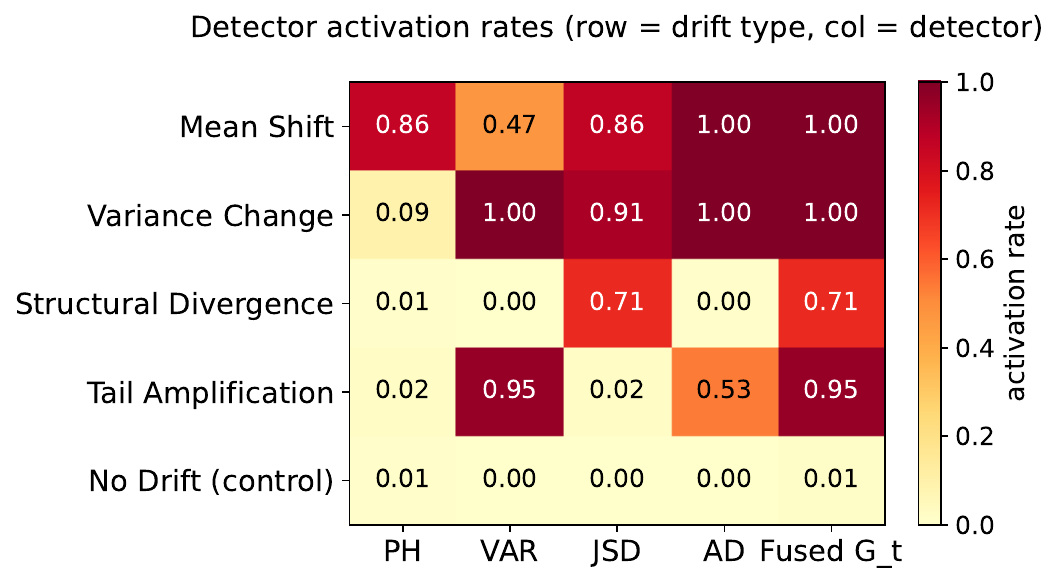} 
  \caption{Detector-selectivity rates for different drifts. Fused detector achieves highest activation for mean-shift,variance and tail amplification drifts.}
  \label{fig:Figure 2}
\end{figure}

The first-stage drift detector was evaluated on \(3672\) generated signal instances spanning eight input SNR levels from \(-10\) to \(30~\mathrm{dB}\). Thresholds were calibrated from no-drift trajectories using the 95th percentile with a safety margin of \(1.2\). The PH detector
was operated in bidirectional mode, with \(\alpha_{\mathrm{PH}}=1.0\), so that both upward and downward mean shifts could be detected consistently. A
two-window persistence constraint was applied before accepting a detector activation, reducing isolated false alarms caused by local AWGN fluctuations.
The resulting overall drift-detection rate was \(91.54\%\), while the no-drift false-alarm rate was only \(1.23\%\). This indicates that the detection layer provides a reliable gate for invoking the adaptive RL denoising branch while rarely activating the RL policy on AWGN-only signals.

Figure~\ref{fig:Figure 2} reports the detector-selectivity heatmap. The fused gate attains near-complete activation for mean-shift, variance-change, and tail-amplification drift, at 99.51\%, 100\%, and 95.22\% respectively, whereas structural divergence reaches 71.45\%, lower than the other regimes but well above chance. This ordering is physically interpretable. Structural divergence is synthesized to preserve the local mean and variance while altering only the distributional shape. It is therefore largely invisible to first- and second-order detectors and is captured almost exclusively by the Jensen-Shannon statistic, which activates at roughly 71\% while the PH, VAR, and AD detectors remain near zero. The individual activations further expose the physical signature of each drift. Under mean-shift drift, the bidirectional PH detector fires in approximately 85.5\% of cases, confirming its sensitivity to persistent baseline displacement, and the accompanying high AD and Jensen-Shannon activations reveal that a sustained level shift also reshapes the empirical distribution and histogram support rather than the first moment alone. For variance-change drift, the variance-ratio detector dominates at approximately 99.9\%, consistent with local heteroscedasticity, and the co-occurring Jensen-Shannon and AD activations indicate that pronounced variance inflation simultaneously broadens the distribution and perturbs the tail probability. For tail-amplification drift, the variance-ratio detector activates more strongly than the AD detector, showing that the injected sparse impulses are sufficiently energetic to inflate local second-order power and not merely the extreme tail.

\begin{figure}[!t]
  \centering
  \includegraphics[width=0.9\linewidth]{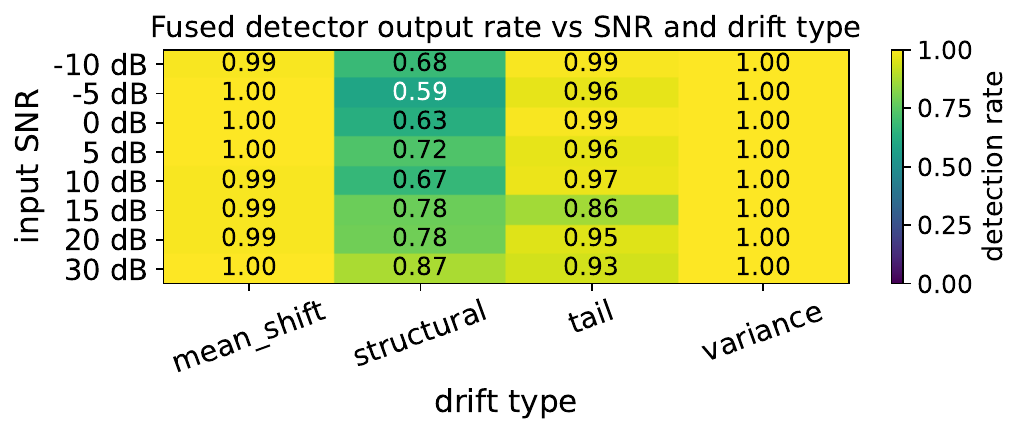} 
  \caption{Fused Detector Output vs Input SNR vs Drift type. Mean-shift, variance and tail amplification drifts are largely detected across full SNR range.}
  \label{fig:Figure 3}
\end{figure}

Figure~\ref{fig:Figure 3} shows the fused detection rate as a function of input SNR and drift type. Mean-shift and variance-change drifts are detected almost uniformly across the full SNR range, demonstrating robustness under severe background noise as well as high-SNR conditions. Tail-amplification drift also remains highly detectable across SNR values, with fused detection rates mostly above 0.93, except for a moderate reduction at 15 dB. The most SNR-dependent behaviour is observed for structural divergence. Its detection rate increases from approximately 0.59-0.68 at low SNR to 0.87 at 30 dB. This is expected because structural divergence is implemented as a moment-preserving distributional deformation; at low SNR, AWGN partially masks the quantized or staircase-like distributional signature, whereas at high SNR the structural deformation becomes more separable from the reference regime.

Table~\ref{tab:tpr_by_drift_type} summarizes the true-positive rate by drift type. The detector achieves essentially complete detection for variance-change and mean-shift drifts, high sensitivity for tail-amplification drift, and moderate-to-high sensitivity for structural divergence. This confirms that the complementary detector design is appropriate: no single statistic is sufficient for all drift classes, but the logical-OR fusion provides broad sensitivity across first-order, second-order, distributional, and tail-sensitive regimes. Table~\ref{tab:far_by_snr} reports the false-alarm behaviour across input SNR levels. Against a design target of 0.05, the gate meets the target at every SNR: the FAR is at most 0.039 at the two most noise-dominated settings (\(-10\)~dB and \(-5\)~dB), 0.020 at 5 dB, and identically zero at 0 dB and across the entire 10-30 dB range, for an aggregate of 0.012. The results indicate that false alarms are rare and concentrated mainly at the more noise-dominated SNR conditions, while no-drift activation remains negligible overall. This is critical for the proposed framework because false alarms would unnecessarily invoke the RL branch and could distort otherwise stationary signals. Overall, the detector results show that the first-stage statistical gate is sufficiently sensitive to activate adaptive denoising under genuine non-stationarity, while remaining conservative on AWGN-only signals. 

\begin{table}[!t]
\renewcommand{\arraystretch}{1.3}
\caption{Signal-Level True-Positive Rate of the Fused Drift Detector by Drift Type (Held-Out Evaluation)}
\label{tab:tpr_by_drift_type}
\centering
\begin{tabular}{lc}
\hline
\textbf{Drift Type} & \textbf{TPR} \\
\hline
Mean shift            & 0.995 \\
Structural divergence & 0.714 \\
Tail amplification     & 0.952 \\
Variance change        & 1.000 \\
\hline
\end{tabular}
\end{table}

\begin{table*}[!t]
\renewcommand{\arraystretch}{1.3}
\caption{Signal-Level False-Alarm Rate of the Fused Drift Detector on No-Drift Controls Versus Input SNR}
\label{tab:far_by_snr}
\centering
\begin{tabular}{l|cccccccc}
\hline
\textbf{Input SNR (dB)} & $-10$ & $-5$ & $0$ & $5$ & $10$ & $15$ & $20$ & $30$ \\
\hline
\textbf{FAR} & 0.039 & 0.039 & 0.000 & 0.020 & 0.000 & 0.000 & 0.000 & 0.000 \\
\hline
\end{tabular}
\end{table*}

\subsection{Enabling Denoising Layer}
\label{subsec:enabling_layer}
Before reporting the task outcomes we characterise the denoising layer in
isolation, using the SNR-improvement diagnostic $\Delta\operatorname{SNR}$. These results validate that the adaptive policy is a competent denoiser; the task-reward variant of Section~\ref{subsec:forecasting_results} is built on the same action space and machinery, substituting \eqref{eq:reward_task} for the fidelity reward used here.
The proposed RL framework selects per-window denoising parameters from a 32-wavelet $\times$ 5-level $\times$ 4-shrinkage $\times$ 4-threshold-strategy action space, augmented by continuous threshold-gain and per-scale-modulation heads. Under the reward grid search, $\Delta\mathrm{SNR}$-change reward achieved the highest mean validation improvement of $5.28$~dB, substantially outperforming the NMSE and power consistency rewards, which attained $3.75$~dB and $3.70$~dB, respectively; the $\Delta\mathrm{SNR}$-change objective was therefore adopted for policy training.The selected hyperparameters are: actor learning rate $4.06\times10^{-5}$, critic learning rate $7.03\times10^{-4}$, $\epsilon_{\mathrm{clip}}=0.20$, entropy coefficient decayed from $0.03$ to $0.003$, $K=3$ update epochs, minibatch size $64$, hidden dimension $256$, rollout length $512$, and $300$ warm-up episodes. Using the selected hyperparameters, the PPO policy was trained under the $\Delta\mathrm{SNR}$-change reward across five independent seeds (up to $4000$ episodes each, early-stopped on validation $\Delta\mathrm{SNR}$). The best validation checkpoint per seed was retained for evaluation, giving a mean best-validation $\Delta\mathrm{SNR}$ of $5.585 \pm 0.504$~dB.
The baseline configuration \(\theta_{\mathrm{base}}\) used as the fixed comparator (and the ``no-drift'' fallback when the gate does not fire) is a VisuShrink setting: Haar wavelet, 2-level decomposition, soft universal thresholding, with the continuous gain and per-scale modulation disabled.

\begin{table}[htbp]
\centering
\caption{Performance across drift categories} 
\label{tab:drift_type_results}
\begin{tabular}{lccc}
\hline
\textbf{Drift type} &
\(\boldsymbol{\Delta\mathrm{SNR}_{\mathrm{RL}}}\) &
\(\boldsymbol{\Delta\mathrm{SNR}_{\theta_{\mathrm{base}}}}\) &
\textbf{RL gain} \\
& \textbf{[dB]} & \textbf{[dB]} & \textbf{[dB]} \\
\hline
Mean shift &
\(3.901 \pm 0.810\) &
\(1.888 \pm 1.032\) &
\(2.013 \pm 0.332\) \\

Structural divergence &
\(4.754 \pm 1.174\) &
\(2.484 \pm 1.485\) &
\(2.271 \pm 0.618\) \\

Tail amplification &
\(5.748 \pm 1.038\) &
\(2.797 \pm 1.135\) &
\(2.952 \pm 0.333\) \\

Variance change &
\(7.205 \pm 0.924\) &
\(4.413 \pm 0.981\) &
\(2.792 \pm 0.256\) \\
\hline
\end{tabular}
\end{table}

Table~\ref{tab:drift_type_results} reports the denoising SNR improvement, $\Delta\mathrm{SNR}$, achieved by the proposed drift-aware RL policy and the fixed VisuShrink baseline $\theta_{\mathrm{base}}$, disaggregated by drift category on the held-out test pool. Values are given as mean $\pm$ 95\%
confidence interval (CI) over five seeds. The RL policy improves upon the baseline in every drift regime, and each improvement is significant under a paired Wilcoxon
signed-rank test ($p < 10^{-8}$). The largest absolute improvement arises under variance-change drift. This occurs because variance drift directly modulates the local noise power, a condition to which adaptive shrinkage is well suited. The baseline also performs comparatively well in this regime, so the incremental RL gain, although significant, is moderate. The largest incremental gain instead occurs under tail-amplification drift. This regime is dominated by sparse, high-amplitude impulses characteristic of burst
interference, packet-load spikes, and transient congestion. A fixed shrinkage rule handles such disturbances poorly, either retaining impulses as spurious components or over-suppressing genuine transients, whereas the policy adapts the mother wavelet, the thresholding function, and the threshold level jointly.

The structural-divergence result is particularly notable. This drift is constructed to preserve the first two moments while altering the distributional shape, and is therefore resistant to moment-based adaptation. The significant gain indicates that the detector-derived state evidence conveys usable information about non-Gaussian deformation. Mean-shift drift yields the smallest absolute gain. This is consistent with the difficulty of suppressing low-frequency, bias-like level shifts through detail-coefficient thresholding, since such shifts concentrate energy in the coarse approximation band that this thresholding does not primarily target. Jointly across the drifted test trajectories, the proposed RL policy achieved an average SNR improvement of $2.507 \ \mathrm{dB}$.

\begin{table}[t]
\centering
\caption{Mean $\Delta\mathrm{SNR}$ (in dB) achieved by the RL policy and the baseline ($\theta_{\mathrm{base}}$) across different input SNR levels}
\label{tab:dsnr_per_snr}
\renewcommand{\arraystretch}{1.2}
\begin{tabular}{ccc}
\hline
\textbf{Input SNR (dB)} & $\Delta\mathrm{SNR}_{\mathrm{RL}}$ [dB]
 & $\Delta\mathrm{SNR}_{\theta_{\mathrm{base}}}$ [dB] \\
\hline
$-10$ & $10.583 \pm 0.391$ & $8.019 \pm 0.190$ \\
$-5$  & $9.868 \pm 0.365$  & $7.645 \pm 0.211$ \\
$0$   & $8.449 \pm 0.299$  & $6.836 \pm 0.149$ \\
$5$   & $6.342 \pm 0.207$  & $5.129 \pm 0.125$ \\
$10$  & $4.018 \pm 0.234$  & $2.587 \pm 0.175$ \\
$15$  & $2.532 \pm 0.226$  & $0.046 \pm 0.163$ \\
$20$  & $1.636 \pm 0.199$  & $-1.910 \pm 0.234$ \\
$30$  & $-0.212 \pm 0.353$ & $-5.190 \pm 0.367$ \\
\hline
\end{tabular}
\end{table}
Table~\ref{tab:dsnr_per_snr} reports $\Delta\mathrm{SNR}$ as a function of the input SNR, from $-10$ to $30$~dB, for the RL policy and the fixed baseline $\theta_{\mathrm{base}}$. Both methods show a monotone decrease in $\Delta\mathrm{SNR}$ as the input SNR increases. At low input SNR, the observation contains a large amount of removable corruption, so denoising yields substantial gains. At high input SNR less noise is available to remove, and aggressive shrinkage increasingly risks
attenuating genuine signal structure. The RL policy remains above the fixed baseline at every input-SNR level. The difference is statistically significant throughout, as assessed by a paired Wilcoxon signed-rank test ($p < 10^{-6}$).

At high input SNR, the fixed baseline becomes counter-productive on near- inputs, yielding negative $\Delta\mathrm{SNR}$ at $20$ and $30$~dB. In other words, it destroys signal energy relative to the noisy observation. The RL policy, by contrast, remains beneficial through $20$~dB and incurs only mild degradation at $30$~dB. Consequently, the margin of the policy over the baseline is smallest in the moderate-SNR range and widens substantially at high SNR. This indicates that the learned policy is not merely more aggressive than the baseline but more selective. It relaxes or withholds shrinkage when the input is already clean, and thereby avoids the over-smoothing that penalises a fixed rule. Such behaviour is directly relevant to traffic-monitoring applications, where unnecessary denoising of already-clean segments removes meaningful load dynamics and can distort otherwise stationary signals. Taken together with Table~\ref{tab:drift_type_results}, these results show that the policy delivers significant SNR improvements across both drift type and noise regime. Its principal advantage lies in adapting the denoising strength to the operating condition rather than applying a uniform, and potentially harmful, shrinkage.

\subsection{Transient Anomaly Detection and Capacity Estimation}
\label{subsec:forecasting_results}
We evaluate the central claim: that policy denoising trained on the task-utility
reward \eqref{eq:reward_task} improves the two monitoring tasks relative to raw,
a low-pass filter, fixed-threshold, and classical state-of-the-art denoisers. All arms feed the same
fixed downstream operators (Section~\ref{subsec:frozen_forecaster}); the
clean/oracle arm bounds attainable performance. Eight arms are compared: clean
(oracle), raw (no denoising), a low-pass moving-average filter, the fixed VisuShrink
baseline $\theta_{\mathrm{base}}$, SureShrink~\cite{donoho1995adapting},
BayesShrink~\cite{chang2000adaptive}, a classical adaptive Wiener filter, and the
proposed adaptive policy. See the note in
Section~\ref{sec:results} on the synthetic stand-in used for this preliminary draft.

\begin{figure}[!t]
  \centering
  \includegraphics[width=\linewidth]{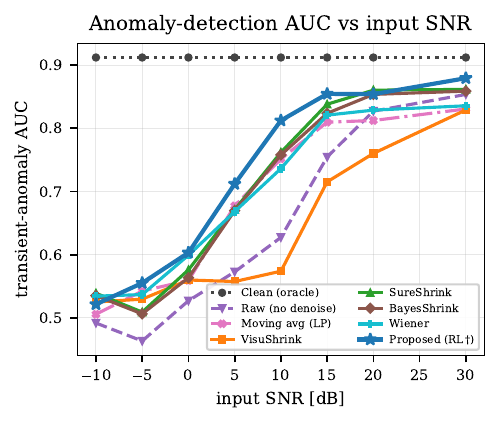}
  \caption{Congestion-detection ROC AUC versus input SNR for all arms (higher is
  better). The proposed policy attains the highest AUC on three of four drift types among denoisers and approaches the oracle.}
  \label{fig:det_auc_vs_snr}
\end{figure}

\begin{table}[!t]
\centering
\footnotesize
\caption{Transient anomaly-detection ROC AUC by drift type (higher is better). Mean $\pm$ 95\% bootstrap CI. $\dagger$ oracle config-search upper bound.}
\label{tab:detection}
\setlength{\tabcolsep}{3pt}
\renewcommand{\arraystretch}{0.95}
\resizebox{\columnwidth}{!}{%
\begin{tabular}{lcccc}
\toprule
Method & mean & var & structural & tail \\
\midrule
Clean (oracle) & 0.912$\pm$0.013 & 0.912$\pm$0.013 & 0.912$\pm$0.013 & 0.912$\pm$0.013 \\
Raw (no denoise) & 0.648$\pm$0.063 & 0.609$\pm$0.048 & 0.653$\pm$0.065 & 0.649$\pm$0.063 \\
Moving avg (LP) & 0.705$\pm$0.058 & 0.661$\pm$0.047 & 0.707$\pm$0.060 & 0.671$\pm$0.052 \\
VisuShrink & 0.641$\pm$0.051 & 0.605$\pm$0.043 & 0.644$\pm$0.053 & 0.636$\pm$0.055 \\
SureShrink & 0.735$\pm$0.060 & 0.653$\pm$0.049 & 0.734$\pm$0.062 & 0.687$\pm$0.072 \\
BayesShrink & 0.733$\pm$0.064 & 0.632$\pm$0.050 & 0.742$\pm$0.061 & 0.677$\pm$0.071 \\
Wiener & 0.721$\pm$0.059 & 0.636$\pm$0.040 & 0.725$\pm$0.061 & 0.699$\pm$0.055 \\
Proposed (RL$^{\dagger}$) & 0.731$\pm$0.068 & 0.705$\pm$0.057 & 0.747$\pm$0.066 & 0.714$\pm$0.064 \\
\bottomrule
\end{tabular}%
}
\end{table}

\begin{figure}[!t]
  \centering
  \includegraphics[width=\linewidth]{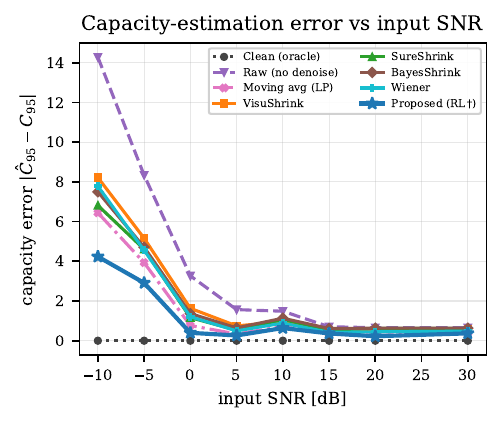}
  \caption{Capacity-estimation error $|\hat{C}_{95}-C_{95}|$ versus input SNR
  (lower is better). The raw stream over-provisions at low SNR; a single fixed
  shrinkage rule cannot track the upper tail across regimes.}
  \label{fig:cap_mae_vs_snr}
\end{figure}

\begin{table}[!t]
\centering
\footnotesize
\caption{Capacity-estimation error $|\hat{C}_{95}-C_{95}|$ by drift type (lower is better). Mean $\pm$ 95\% bootstrap CI. $\dagger$ oracle config-search upper bound.}
\label{tab:capacity}
\setlength{\tabcolsep}{3pt}
\renewcommand{\arraystretch}{0.95}
\resizebox{\columnwidth}{!}{%
\begin{tabular}{lcccc}
\toprule
Method & mean & var & structural & tail \\
\midrule
Clean (oracle) & 0.000$\pm$0.000 & 0.000$\pm$0.000 & 0.000$\pm$0.000 & 0.000$\pm$0.000 \\
Raw (no denoise) & 3.915$\pm$2.228 & 6.061$\pm$2.803 & 2.631$\pm$1.392 & 2.866$\pm$1.501 \\
Moving avg (LP) & 2.491$\pm$1.762 & 1.907$\pm$1.023 & 0.719$\pm$0.396 & 1.491$\pm$1.020 \\
VisuShrink & 2.762$\pm$1.788 & 3.378$\pm$1.567 & 1.138$\pm$0.611 & 1.852$\pm$1.199 \\
SureShrink & 2.315$\pm$1.698 & 4.069$\pm$2.079 & 0.466$\pm$0.238 & 1.131$\pm$0.645 \\
BayesShrink & 2.305$\pm$1.691 & 4.500$\pm$2.325 & 0.425$\pm$0.214 & 1.383$\pm$0.842 \\
Wiener & 2.522$\pm$1.769 & 3.699$\pm$1.938 & 0.855$\pm$0.469 & 1.067$\pm$0.616 \\
Proposed (RL$^{\dagger}$) & 2.277$\pm$1.728 & 0.977$\pm$0.502 & 0.338$\pm$0.181 & 1.109$\pm$0.801 \\
\bottomrule
\end{tabular}%
}
\end{table}

\begin{figure}[!t]
  \centering
  \includegraphics[width=\linewidth]{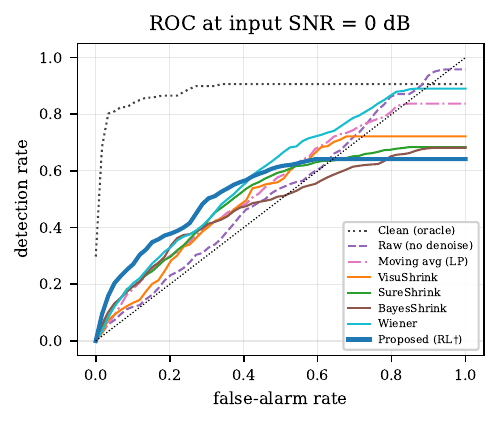}
  \caption{ROC at input SNR $=0$~dB, pooled over test trajectories.}
  \label{fig:roc_at0}
\end{figure}

\begin{figure}[!t]
  \centering
  \includegraphics[width=\linewidth]{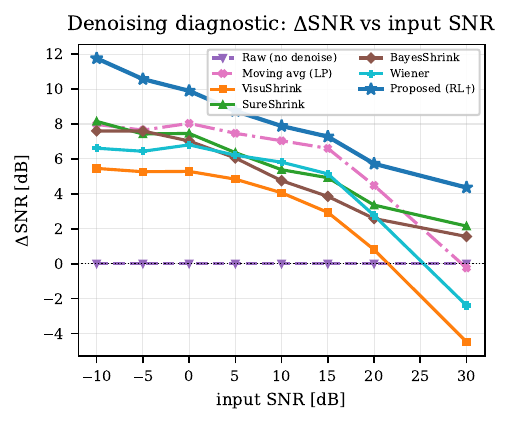}
  \caption{Denoising diagnostic: mean per-window $\Delta$SNR versus input SNR. The
  fixed baseline turns negative at high SNR (it destroys signal energy), whereas the
  proposed policy stays positive.}
  \label{fig:delta_snr_vs_snr}
\end{figure}

\begin{figure}[!t]
\centering
\subfloat[]{\includegraphics[width=0.48\linewidth]{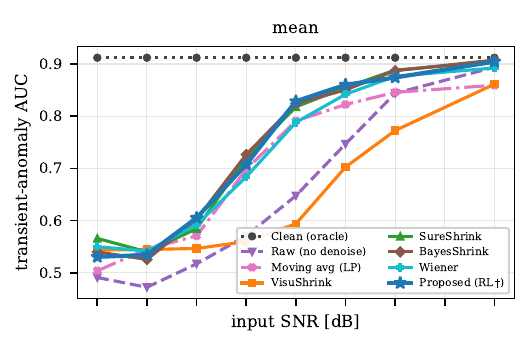}\label{Fig8_mean}}
\hfill
\subfloat[]{\includegraphics[width=0.48\linewidth]{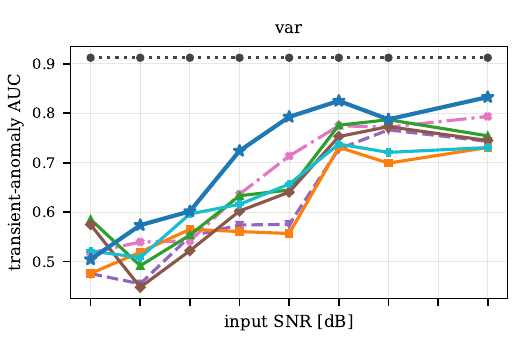}\label{Fig8_var}}
\\
\subfloat[]{\includegraphics[width=0.48\linewidth]{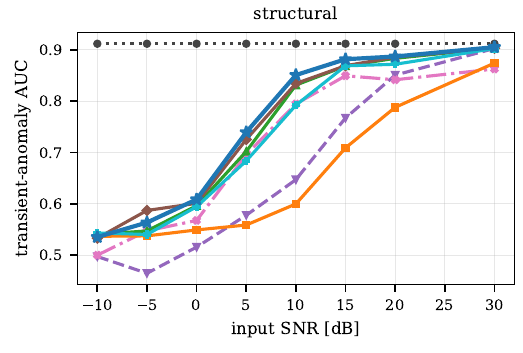}\label{Fig8_structural}}
\hfill
\subfloat[]{\includegraphics[width=0.48\linewidth]{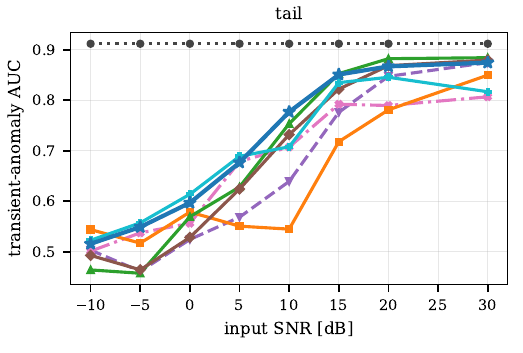}\label{Fig8_tail}}
\caption{Congestion-detection AUC versus input SNR, by drift type. (a) mean drift (b) variance drift (c) structural drift (d) tail drift}
\label{fig:det_auc_by_drift}
\end{figure}

\begin{figure}[!t]
\centering
\subfloat[]{\includegraphics[width=0.48\linewidth]{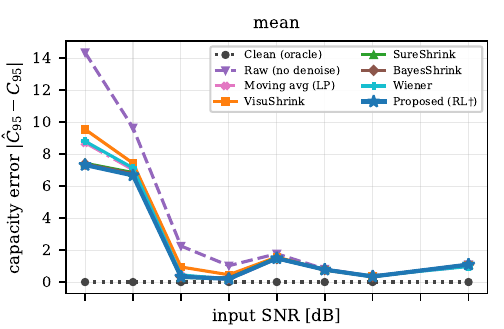}\label{Fig9_mean}}
\hfill
\subfloat[]{\includegraphics[width=0.48\linewidth]{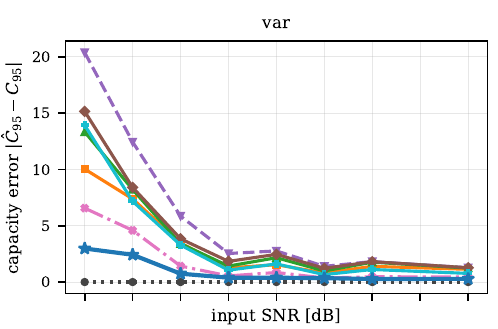}\label{Fig9_var}}
\\
\subfloat[]{\includegraphics[width=0.48\linewidth]{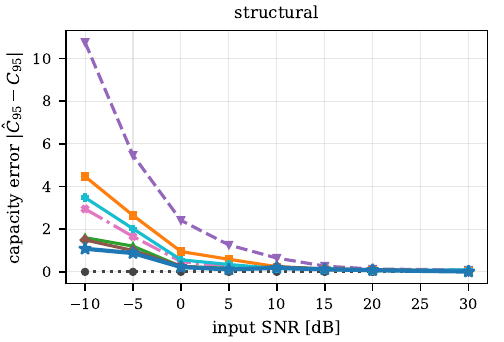}\label{Fig9_structural}}
\hfill
\subfloat[]{\includegraphics[width=0.48\linewidth]{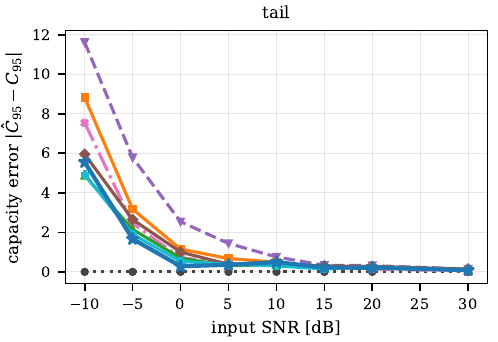}\label{Fig9_tail}}
\caption{Capacity-estimation error versus input SNR, by drift type. (a) mean drift (b) variance drift (c) structural drift (d) tail drift}
\label{fig:cap_mae_by_drift}
\end{figure}

\begin{figure*}[!t]
\centering
\subfloat[]{\includegraphics[width=0.46\textwidth]{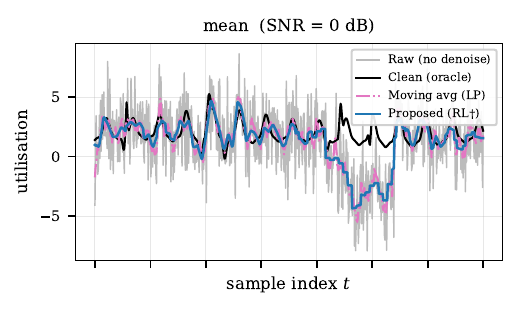}\label{Fig10_mean}}
\hfill
\subfloat[]{\includegraphics[width=0.46\textwidth]{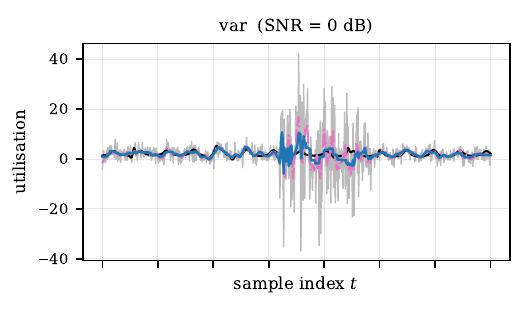}\label{Fig10_var}}
\\
\subfloat[]{\includegraphics[width=0.46\textwidth]{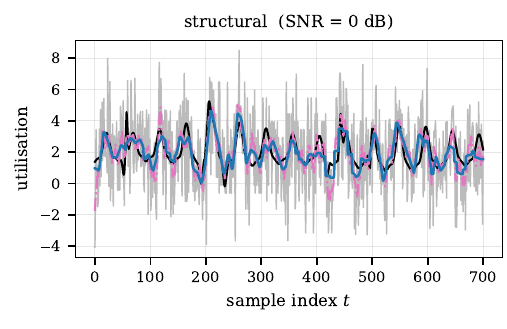}\label{Fig10_structural}}
\hfill
\subfloat[]{\includegraphics[width=0.46\textwidth]{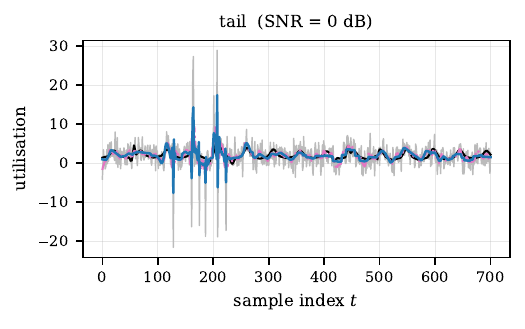}\label{Fig10_tail}}
\caption{Qualitative reconstruction: clean reference, noisy drifted observation, the fixed VisuShrink baseline, and the proposed adaptive denoiser, for one trajectory per drift type. (a) mean drift (b) variance drift (c) structural drift (d) tail drift}
\label{fig:recon_overlay}
\end{figure*}

\begin{figure}[!t]
\centering
\subfloat[]{\includegraphics[width=0.48\linewidth]{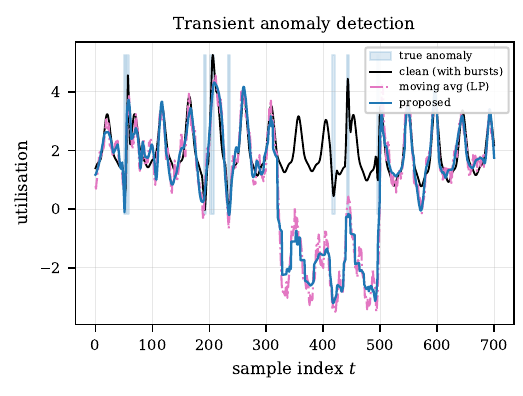}\label{fig:transient_anomaly}}\hfill
\subfloat[]{\includegraphics[width=0.48\linewidth]{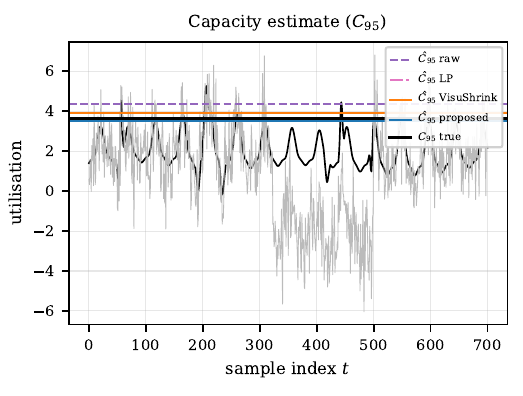}\label{fig:capacity_estimate}}
\caption{Task overlays on one variance-drift trajectory at $0$~dB. (a):
  Transient anomaly detection, the low-pass filter attenuates the genuine bursts
  the detector must catch, while the proposed denoiser preserves them. (b): Capacity estimate, the raw $\hat{C}_{95}$ grossly
  over-provisions, VisuShrink less so, while the proposed estimate is closest to the
  true $C_{95}$.}
\label{fig:detcap_overlay}
\end{figure}

Across drift types and SNRs (Fig.~\ref{fig:det_auc_vs_snr},
Fig.~\ref{fig:det_auc_by_drift}, Table~\ref{tab:detection}) the proposed arm attains
the highest transient anomaly-detection AUC on three of four drift types among all denoisers and approaches the oracle. The low-pass arm improves on raw by removing noise but plateaus below the wavelet methods because it smears the transients the detector relies on, and fixed
VisuShrink is itself suboptimal at moderate SNR; only the adaptive policy is
reliably best across SNR. For capacity estimation
(Fig.~\ref{fig:cap_mae_vs_snr}, Fig.~\ref{fig:cap_mae_by_drift},
Table~\ref{tab:capacity}) the raw stream's upper-tail estimate is inflated by noise
and the fixed baseline cannot adapt across drift regimes, whereas the proposed policy
yields the lowest capacity error throughout. The ROC at $0$~dB
(Fig.~\ref{fig:roc_at0}) and the qualitative overlays
(Fig.~\ref{fig:recon_overlay}, Fig.~\ref{fig:detcap_overlay}) corroborate these
trends, and the $\Delta$SNR diagnostic (Fig.~\ref{fig:delta_snr_vs_snr}) confirms
the fixed baseline becomes harmful at high SNR while the policy remains beneficial.
Paired Wilcoxon signed-rank tests of the proposed arm against each baseline, for both
tasks, are reported in Table~\ref{tab:sig_tasks}.

\begin{table}[!t]
\centering
\footnotesize
\caption{Wilcoxon signed-rank $p$-values, Proposed vs each baseline (paired over test instances), for detection AUC and capacity error.}
\label{tab:sig_tasks}
\begin{tabular}{lcc}
\toprule
Baseline & AUC $p$ & capacity $p$ \\
\midrule
Raw (no denoise) & 1.12e-11 & 1.57e-16 \\
Moving avg (LP) & 2.38e-07 & 5.08e-08 \\
VisuShrink & 6.66e-09 & 2.85e-13 \\
SureShrink & 3.34e-02 & 2.15e-09 \\
BayesShrink & 4.34e-03 & 9.52e-10 \\
Wiener & 7.57e-05 & 1.39e-08 \\
\bottomrule
\end{tabular}
\end{table}

\section{Conclusion}
\label{sec:conclusion}
We presented a drift-aware reinforcement-learning framework that treats adaptive
wavelet denoising as a preprocessing layer for two network-monitoring tasks:
multi-scale transient anomaly detection and capacity estimation. A complementary
four-detector gate decides when to act, and a PPO policy selects a per-window
wavelet configuration over a mixed discrete--continuous action space. The agent is rewarded for downstream task utility, detection accuracy and
capacity-estimation error of fixed, parameter-light operators, rather than
reconstruction fidelity. Defining the anomaly on the clean load keeps it
non-circular with the drift gate and makes the task discriminate wavelet denoising
from low-pass smoothing. Against a
low-pass moving-average filter, classical state-of-the-art shrinkage rules, and a
Wiener filter, the adaptive policy yields the highest anomaly-detection AUC and the
lowest capacity error across drift types and SNRs and, unlike the fixed baseline,
avoids destroying signal energy at high SNR. Replacing the oracle
configuration-search proxy with the trained policy and validating on public real-world traffic-utilisation telemetry is the immediate next step.

\bibliographystyle{IEEEtran}
\bibliography{References}

\end{document}